\documentclass[prl,twocolumn,nofootinbib,aps,tightenlines,preprintnumbers,notitlepage,longbibliography,superscriptaddress]{revtex4-1}
\usepackage{graphicx}
\usepackage{amsmath}
\usepackage{amsfonts}
\usepackage[colorlinks=true,citecolor=blue,urlcolor=cyan,linkcolor=black]{hyperref}

\usepackage{amsmath}
\usepackage{amsfonts}

\usepackage{tabularx}
\newcolumntype{C}{>{\centering\arraybackslash}X}
\newcolumntype{R}{>{\raggedleft\arraybackslash}X}

\usepackage[usenames, dvipsnames]{color}
\usepackage[normalem]{ulem}
\usepackage{xcolor}

\newcommand{\dd}{{\rm d}}

\newcommand{\bk}{\boldsymbol{k}}
\newcommand{\be}{\begin{eqnarray}}

\newcommand{\ee}{\end{eqnarray}}

\definecolor{colorA}{HTML}{1E90FF}
\definecolor{colorB}{HTML}{228B22}
\definecolor{colorC}{HTML}{FF7F00}
\definecolor{colorD}{HTML}{4B0082}
\definecolor{colorE}{HTML}{B22222}

\definecolor{lgreen}{HTML}{32CD32}
\definecolor{lgray}{HTML}{D3D3D3}
\definecolor{dblue}{HTML}{1E90FF}
\definecolor{dblue}{HTML}{1E90FF}
\definecolor{orange}{HTML}{FF4500}
\definecolor{indigo}{HTML}{4B0082}

\definecolor{teal}{HTML}{008080}
\definecolor{firebrick}{HTML}{B22222}
\definecolor{salmon}{HTML}{FA8072}
\definecolor{darkgreen}{HTML}{006400}

\newcommand{\jhu}{William H. Miller III Department of Physics and Astronomy, Johns Hopkins University, Baltimore, MD 21218, USA}

\newcommand{\illa}{Astronomy Department, University of Illinois at Urbana-Champaign, 1002 W. Green Street, Urbana, IL 61801, USA}

\newcommand{\illb}{Department of Physics, University of Illinois Urbana-Champaign, 1110 W. Green Street, Urbana, IL 61801, USA}

\newcommand{\perimeter}{Perimeter Institute for Theoretical Physics, 31 Caroline St N, Waterloo, ON N2L 2Y5, Canada}

\newcommand{\york}{Department of Physics and Astronomy, York University, Toronto, ON M3J 1P3, Canada}

\newcommand{\berkeleya}{Lawrence Berkeley National Laboratory, One Cyclotron Road, Berkeley, CA 94720, USA}
\newcommand{\berkeleyb}{Berkeley Center for Cosmological Physics, Department of Physics, University of California, Berkeley, CA 94720, USA}

\graphicspath{ {plots/} }

\begin{document}

\title{Probing helium reionization with kinetic Sunyaev Zel'dovich tomography}

\author{Selim~C.~Hotinli}
\affiliation{\jhu}

\author{Simone~Ferraro}
\affiliation{\berkeleya}
\affiliation{\berkeleyb}

\author{Gilbert~P.~Holder}
\affiliation{\illa}
\affiliation{\illb}

\author{Matthew~C.~Johnson}
\affiliation{\perimeter}
\affiliation{\york}

\author{Marc~Kamionkowski}
\affiliation{\jhu}

\author{Paul~La~Plante}
\affiliation{\berkeleyb}

\begin{abstract}

{Reionization of helium is expected to occur at redshifts $z\sim3$ and have important consequences for quasar populations, galaxy formation, and the morphology of the intergalactic medium, but there is little known empirically about the process.  Here we show that kinetic Sunyaev-Zeldovich (kSZ) tomography, based on the combination of CMB measurements and galaxy surveys, can be used to infer the primordial helium abundance as well as the time and duration of helium reionization. {We find a high-significance detection at  ${\sim10\sigma}$ can be expected from Vera Rubin Observatory and CMB-S4 in the near future.} A more robust  characterization of helium reionization will require next-generation experiments like MegaMapper (a proposed successor to DESI) and CMB-HD.}

\end{abstract}

\maketitle

Probing helium reionization\footnote{Note that throughout this work we refer to the ionization of the second electron of helium as the helium reionization.}---one of the major large-scale transitions of the intergalactic medium (IGM)--- has great potential significance for understanding the formation of galaxies and quasar activity at early times, and may open a new window on big bang nucleosynthesis. Since photons emitted by the first stars (sourcing the reionization of hydrogen) are not energetic enough\footnote{The ionization energy of the second electron in helium is 54.4eV, while the ionization energy of hydrogen is 13.6eV.} to fully ionize helium, helium reionization occurs only after the emergence of a substantial number of quasars. As a result, the history of helium reionization strongly depends on the properties of quasars, such as their luminosity function~\citep{Ross2013, Masters2012, McGreer2013, McGreer2018, Pan2022}, accretion mechanisms and other astrophysics~\citep{Shen:2014rka}, clustering, variability, lifetimes~\citep{Hopkins:2006vv, Schmidt2017}, as well as the general growth and evolution of super-massive black holes~\citep{Inayoshi:2019fun}. Since essentially all of the helium in the Universe is ultimately doubly ionized, the total change in the ionization fraction is a measure of the primordial helium abundance---a sensitive probe of big bang nucleosynthesis. Probing helium reionization can also improve our understanding of relativistic species through improving the primordial helium fraction $Y_p$ measurement and breaking the degeneracy between number of relativistic degrees of freedom $N_{\rm eff}$ and $Y_p$. The primordial helium abundance depends on the weak interaction rates as well as the neutron lifetime, and improving its measurement can allow further valuable insights into our cosmological history. 

There is evidence for quasar activity peaking around $z\sim3$ \citep{Richards2006}, which coincides with measurements of the helium Ly$\alpha$ forest suggesting the helium in the IGM has not yet been doubly ionized~\citep{Jakobsen1994,Zheng2008,Syphers2014}. Measurements of the thermal history of the intergalactic medium (IGM) have provided indirect evidence for helium reionization occurring roughly $2.5 \lesssim z \lesssim 4$ \citep{Calura2012,Viel2013,Boera2016}, with semi-numeric and hydrodynamic simulations of helium reionization supporting a similar picture \citep{UptonSanderbeck2016,LaPlante2017,LaPlante2018,Bolton2017}. Nevertheless, the precise details of the timing, duration, and morphology of helium reionization remain largely uncertain. Surveys of the helium Ly$\alpha$ forest are severely limited by intervening Lyman-limit systems at lower redshift \citep{Syphers2012}, which means it will be challenging to make further progress. Furthermore, measurements of the hydrogen Ly$\alpha$ forest are indirect, and do not provide a clear picture of the ionization state of helium.   Additional probes of helium reionization will be incredibly valuable. For example, it has been shown that future large catalogs of fast radio bursts could probe helium reionization \citep{2020PhRvD.101j3019L,2021PhRvD.103j3526B},

In this letter, we propose a new way to detect and characterize helium reionization by means of tomography using the kinetic Sunyaev-Zel'dovich (kSZ) effect \citep{Sunyaev1972}. This kSZ tomography has been shown to be an effective way to extract cosmological information (through the reconstructed radial-velocity field) from small-scale fluctuations in the cosmic microwave background (CMB) and a tracer of the electron density, such as a galaxy survey~\citep[e.g.][]{Deutsch:2017ybc,Smith:2018bpn,Munchmeyer:2018eey,Zhang:2015uta,Hotinli:2019wdp,Cayuso:2019hen,Alvarez:2020gvl,Ferraro:2018izc, Smith:2016lnt, Hotinli:2020csk}.  Ongoing large-scale structure surveys that access $2<z<5$ galaxy and quasar populations such as DESI~\citep{DESI:2016fyo} or Vera Rubin Observatory (VRO)~\citep{2009arXiv0912.0201L} are opening a new window of opportunity into probing the Universe at largely-uncharted epochs of structure formation.   Cross correlation of large-scale structure (LSS) measured at these redshifts with maps of the CMB can be a powerful probe in the near future. We demonstrate that by measuring the statistical variations of the cross correlation between the LSS and CMB, one can probe the change in the mean ionization fraction during the epoch of helium reionization to high significance with upcoming surveys such as CMB-S4~\citep{Abazajian:2016yjj,Abazajian:2019eic}, together with DESI, VRO or the proposed MegaMapper~\citep{Schlegel:2019eqc, Ferraro:2022cmj}.

The CMB temperature anisotropy induced by the kSZ effect from large-scale structure {in a} shell of width $L_{\rm shell}$ at a redshift $z=z_*$ is
\be
\Theta_{\rm kSZ}(\boldsymbol{\theta})=K(z_*)\int^{L_{\rm shell}}_0 \dd r\, q_\parallel(\boldsymbol{r}), 
\ee
where $\boldsymbol{r}\equiv\chi_\star\boldsymbol{\theta}+r\hat{\boldsymbol{r}}$, $\boldsymbol{\theta}$ is the angular direction on the sky, $\chi_\star$ is the conformal distance to the shell, $\hat{\boldsymbol{r}}$ is the unit vector in the radial direction, $\Theta(\boldsymbol{\theta})$ is the fractional fluctuation of CMB temperature, and
\be
K(z)=-\sigma_T n_{\rm H}x_e(z)e^{-\tau(z)}(1+z)^2\,,
\ee 
is the radial weight function in units of Mpc$^{-1}$. Here, $\sigma_T$ is the Thomson scattering cross-section, $\tau(z)$ is the optical depth to redshift $z$, $n_{\rm H}$ is the {hydrogen} number density, $x_e(z)$ is the number of free electrons per hydrogen atom, and $q_\parallel(\boldsymbol{r})=\delta_e(\boldsymbol{r})v_\parallel(\boldsymbol{r})$ is the electron-momentum field, projected onto the radial direction. {The velocity field $v_\parallel(\boldsymbol{r})$ can be reconstructed at cosmological scales from its influence on the correlation between the electron-momentum field and large-scale structure.} The (inverse) noise on the reconstructed velocity is given by~\citep{Smith:2018bpn} 
\be\label{eq:ksz_tomo}
\frac{1}{N_\parallel(\boldsymbol{k}_L)}=\frac{K_*^2}{\chi_*^2}\int\!\!\frac{k_s\dd k_s}{2\pi}\!\left(\!\frac{P_{\rm ge}(k_s)^2}{P^{\rm obs}_{\rm gg}(k_s)C_\ell^{TT,\rm obs}}\!\right)_{\!\!\!\ell=k\chi_*}\!\!\!\!\,,\!\!\,\,\,\,\,\,
\ee
where $\bk$ is the three-dimensional Fourier wavevector and the integral is over small-scale Fourier modes $k_S$. We represent large-scale modes with an `$L$' subscript.  Here, $C_\ell^{TT,\rm obs}$ is the observed CMB spectrum {including foregrounds and noise}, $P^{\rm obs}_{\rm gg}(k)$ is the observed galaxy power spectrum and $P_{\rm ge}(k)$ is the power-spectrum of the galaxy-electron correlation.

On large scales where linear theory is valid, the reconstructed velocity fields are proportional to the cosmic growth rate. The reconstructed velocity amplitude is proportional to the free-electron density at a given redshift and satisfy
\be
\hat{v}_{\parallel}(\bk,z)&=[{\bar{x}_e(z)}/{\bar{x}_e(z)_{\rm fid}}]b_\parallel(z)\,\mu\frac{f a H}{k}\delta_{\rm m}(z,\boldsymbol{k})\,,
\ee
where $\bar{x}_e(z)/{\bar{x}_e(z)}_{\rm fid}$ is equal to unity for a given fiducial cosmology with helium reionization, $b_\parallel(z)$ is the optical-depth bias due to mismodelling of the small-scale electron-galaxy cross-correlation as described in Ref.~\citep{Smith:2018bpn}, $f$ is the linear-theory growth rate, $a$ is the scale factor and $H$ is the Hubble parameter. As a result, the reconstructed velocity fields probe the mean ionization fraction: if the helium reionization is not accounted for, the velocity amplitudes will be biased by the change of the mean reionization fraction. The combination of the galaxy and the velocity satisfies
\be
P_{\rm gg}(k,\mu,z)\!&=&\!(b_g(z)+f\mu^2)^2P_{\rm mm}(k,z)\label{eq:galaxy_spec}\,,\\
P_{vv}(k,\mu,z)\!&=&\!\!\left(\!\frac{\bar{x}_e(z)}{\bar{x}_e(z)_{\rm fid}}\!\right)^2\!\!\!b_\parallel(z)^2\!\left(\!\frac{faH}{k}\!\right)^2\!\!\!P_{\rm mm}(k,z)\,,\\
P_{{\rm g}v}(k,\mu,z)&&\\
=\left(\!\frac{\bar{x}_e(z)}{\bar{x}_e(z)_{\rm fid}}\!\right)&\,&\!\!\!\!\!\!\!b_\parallel(z)\left(\!\frac{faH}{k}\!\right)(b_g(z)\!+\!f\mu^2)P_{\rm mm}(k,z)\,,\nonumber
\ee
where $P_{\rm mm}(a)=D^2(a)P_{\rm mm}(a=1)$ is the matter power-spectra and $D(a)$ is the linear-theory growth factor for the matter spectrum that parameterizes the time evolution of the matter power-spectra, {and $b_g(z)$ is the galaxy bias which relates the matter distribution to the galaxy.}

\begin{figure}[t!]
    \includegraphics[width=1.1\columnwidth]{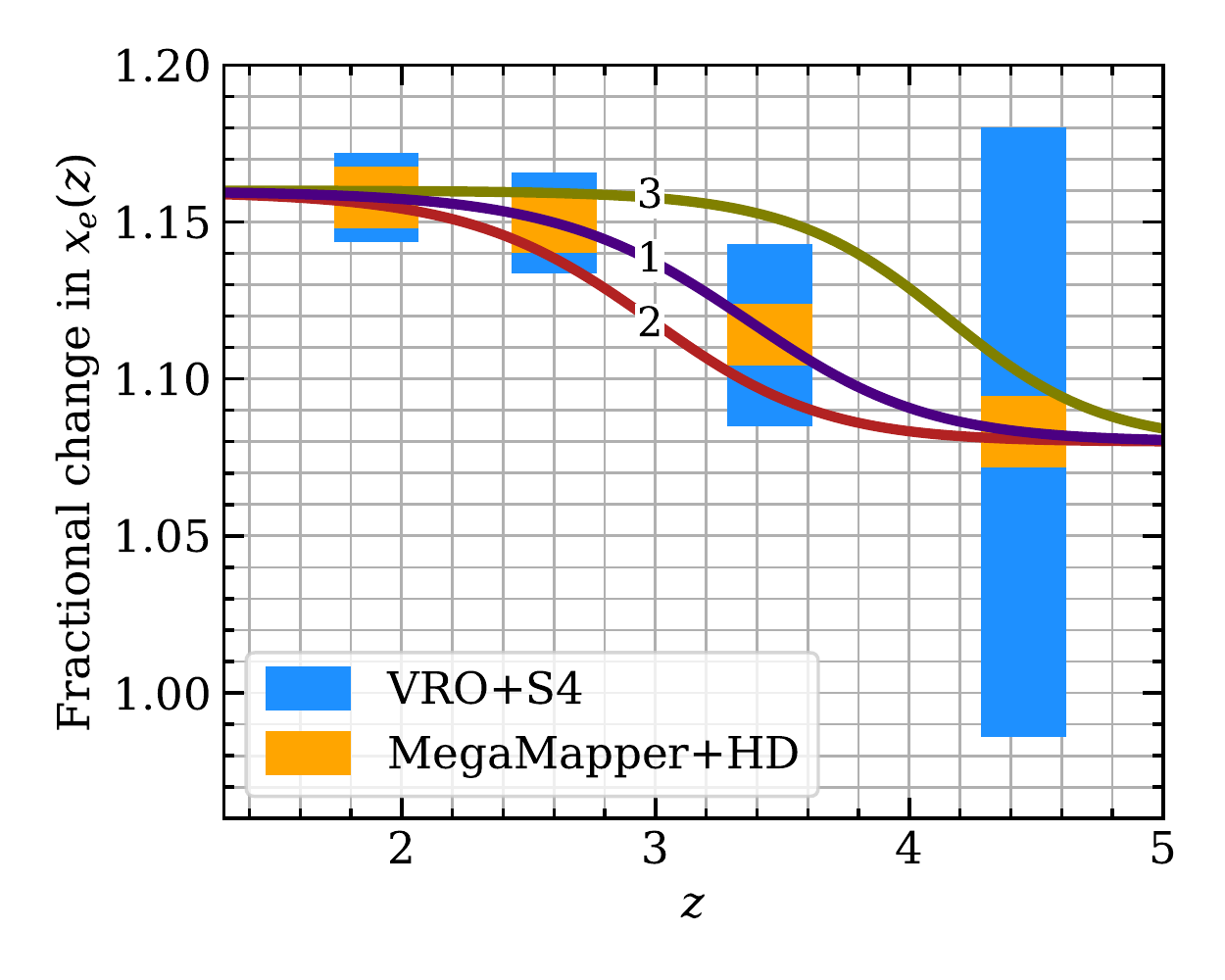}
    \vspace{-1.1cm}
    \caption{Fractional change in the electron fraction $x_e(z)$ during helium reionization of the three models we consider here. The error bars correspond to the measurement accuracy on the optical-depth bias $b_\parallel(z)$, representative of the error on the amplitude of the reconstructed radial velocity, as discussed in the text. Here, we include forecasts for the combination of VRO and CMB-S4, and MegaMapper and CMB-HD.} 
    \vspace*{-0.5cm}
    \label{fig:constraints1}
\end{figure}

We characterise the  change in the ionization fraction during helium reionization with a hyperbolic tangent defined as
\be\label{eq:mean_reio}
\overline{x}_e(z)=\frac{1}{2}\left[2+\Delta \bar{x}_{\rm He}+\Delta \bar{x}_{\rm He}\tanh{\left(\frac{y(z)-y^{\rm He}_{\rm re}}{\Delta_y^{\rm He}}\right)}\right],\,\,\,
\ee
as commonly done in the standard theory codes such as CAMB~\citep{Ali-Haimoud:2010hou}. Here, $y(z)=(1+z)^{3/2}$, $\Delta \bar{x}_{\rm He}$ determines the total change in the mean ionization fraction during helium reionization (or equivalently, mean helium density fraction), $y_{\rm re}^{\rm He}$ sets the redshift of Helium reionization and $\Delta_y^{\rm He}$ parameterizes the duration of the transition. In what follows we will trade $\Delta\bar{x}_{\rm He}$ with $Y_p$ {and use CAMB to calculate $\partial Y_p/\partial\Delta\bar{x}_{\rm He}$}, $y_{\rm re}^{\rm He}$ with the redshift half-way-through the helium reionization we define with $y_{\rm re}^{\rm He}=(1+z_{\rm re}^{\rm He})^{3/2}$ and the $\Delta_y^{\rm He}$ parameter with $\Delta_z^{\rm He}$, which we define as the duration in redshift of the central 50$\%$ change in ionization fraction. In Fig.~\ref{fig:constraints1}, we demonstrate three reionization models labeled with numbers 1 to 3 with fiducial choices for ($z_{\rm re}^{\rm He},\Delta_z^{\rm He}$) set equal to $(3.34,0.8)$, $(2.29,0.79)$ and $(4.14,0.58)$, respectively. We take $Y_p=0.245$ for all models.  These models are chosen to roughly match models H1, H3 and H6, considered in Ref.~\citep{LaPlante:2016bzu}, respectively, and represent several plausible models of helium reionization. Model H1 reproduces the quasar abundance measured by Refs.~\citep{Ross2013,Masters2012,McGreer2013}, the typical quasar spectrum measured by Ref.~\citep{Lusso2015}, and quasar clustering measured by BOSS \citep{White2012}. Model H3 uses a quasar abundance reduced by a factor of 2, which is consistent with the measured uncertainties but yields a slightly later reionization scenario. Model H6 uses a uniform UV background rather than explicit quasar sources, and reproduces the semi-numeric models of Ref.~\citep{Haardt2012}. Distinguishing between these models can provide an independent determination of the average abundance and luminosity of quasars, which complements direct measurements from surveys like SDSS. Quasar activity also significantly heats the IGM, which in turn affects measurements of the low-density gas in the Ly$\alpha$ forest.

\begin{table}[t!]
  \begin{center}
    \begin{tabular}{| l | l l l l |} 
    \hline
     VRO & $\,\!z\!=1.9$ & 2.6 & 3.45 & 4.45  \\  
     $b_g$ & 1.81 &  2.47 & 3.28 & 4.23 \\
     $n_{\rm gal}$ ($\times10^{4}$) & 14.9 &  2.9 & 0.34 & 0.02  \\
     \hline
     MegaMapper & & & &  \\ 
     $b_g$ & 1.92 & 3.18 & 4.71 & 6.51  \\
     $n_{\rm gal}$ ($\times10^{4}$) & 11.7 &  3.4 & 1 & 0.2  \\
    \hline
   \hline
   \hline
   \end{tabular}
   \vspace{-0.2cm}
    \caption{\textit{Assumed galaxy bias $b_g$ and number density $n_{\rm gal}$ at each redshift bin.} {For VRO, we approximate the galaxy density of the ``gold'' sample, with $n(z) = n_\text{gal}[({z}/{z_0}]^2\exp(-z/z_0)/{2z_0}$ with $n_\text{gal}=40~\text{arcmin}^{-2}$ and $z_0=0.3$ and take the galaxy bias as $b_g(z)=0.95/D(z)$. Our calculation of the number density and the galaxy bias of MegaMapper, which is proposed as a follow-up to DESI that will target Lyman-break galaxies (LBGs) and Lyman-alpha emitters and use deeper VRO images, is described in Ref.~\citep{Foreman:2022ksz}, which follows Ref.~\citep{Ferraro:2019uce}, using galaxies with threshold apparent magnitude $m_{\rm UV}^{\rm th}=24.5$ (matching the limiting magnitude assumed for the ``idealised sample'' from Ref.~\citep{Ferraro:2019uce}) and using the ``linear HOD model" fit of Ref.~\citep{Harikane:2017lcw} at $z \simeq 3.8$.} }
    \label{tab:survey-specs}
  \end{center}
  \vspace{-0.8cm}
\end{table}

In order to assess the prospects to detect helium reionization, we consider three LSS surveys; the ongoing measurements of quasi-stellar objects (QSOs) with DESI~\citep{2016arXiv161100036D}, {the photometric VRO survey~\citep{2009arXiv0912.0201L},} and high-$z$ galaxy measurements from the proposed MegaMapper~\citep{Schlegel:2019eqc}. We describe the survey specification of these experiments in Table~\ref{tab:survey-specs}.\footnote{A quick forecast of DESI quasars, calculating the number density following Ref.~\citep{2016arXiv161100036D} and setting the bias to satisfy $b_g(z)=1.2/D(z)$, shows that it would be difficult to detect helium reionization with this data. We drop DESI from our analysis in what follows.} We consider 4 redshift boxes centered at $z\in\{1.9, 2.6, 3.45, 4.45\}$. We assume a sky fraction of $f_{\rm sky}\simeq0.5$ which roughly gives volumes of \{150, 200, 220, 240\}\,Gpc$^3$ at each redshift box, respectively. 

The total CMB power gets contributions from weak gravitational lensing, the kSZ effect (both from reionization and late times), {other foregrounds}, as well as experimental noise satisfying 
\be
N_\ell=\Delta_T^2\exp\left[\frac{\ell(\ell+1)\theta^2_{\rm FWHM}}{8\ln2}\right]\,,
\ee
where we consider {three} CMB experiments with white noise specifications matching Simons Observatory (SO), CMB-S4 and CMB-HD {as given in Table~\ref{tab:beamnoise}}. We also include the frequency-dependent clustered CIB, Poisson CIB and tSZ foregrounds, the black-body late-time and reionization kSZ, and radio sources as described in Ref.~\citep{Hotinli:2021hih}. {We calculate the lensed CMB black-body using CAMB~\cite{CAMB}. Our ILC-cleaning procedure is explained in Ref.~\citep{Hotinli:2021hih}.}

\begin{table}[t!]
\begin{tabular}{|l|c|c|c|c|c|c|}
    \hline & \multicolumn{3}{c|}{Beam FWHM} & \multicolumn{3}{c|}{Noise RMS~$\mu$K'} \\ 
    \cline{2-7} 
    & \multicolumn{1}{c|}{SO} & \multicolumn{1}{c|}{CMB-S4} & \multicolumn{1}{c|}{CMB-HD} & \multicolumn{1}{c|}{SO} & \multicolumn{1}{c|}{CMB-S4} & \multicolumn{1}{c|}{CMB-HD} \\ \hline
    39 GHz  & $5.1'$ & $5.1'$ & $36.3''$ & 36 & 12.4 & 3.4 \\
    93 GHz  & $2.2'$ & $2.2'$ & $15.3''$ & 8 & 2.0 & 0.6 \\
    145 GHz & $1.4'$ & $1.4'$ & $10.0''$ & 10 & 2.0 & 0.6 \\
    225 GHz & $1.0'$ & $1.0'$ & $6.6''$ & 22 & 6.9 & 1.9 \\
    280 GHz & $0.9'$ & $0.9'$ & $5.4''$ & 54 & 16.7 & 4.6 \\ \hline
\end{tabular}
\vspace{-0.2cm}
\caption{{\it Inputs to ILC noise: } The beam and noise RMS parameters we assume for survey configurations roughly corresponding to Simons Observatory (SO) (baseline), CMB-S4 and CMB-HD.}
\vspace{-0.4cm}
\label{tab:beamnoise}
\end{table}

In addition to the three parameters defined above that characterise the helium reionization, we model the galaxy and velocity power spectra with the linear-theory growth rate $f$, the amplitude $\sigma_8$ of matter fluctuations on the scale of $8h^{-1}$Mpc, and independent galaxy and optical-depth bias parameters $b_{g}(z)$ and $b_{\parallel}(z)$ at each redshift. Note that throughout this work, we assume measurements at lower-redshifts will provide $\lesssim1\%$ priors on $f$ and $\sigma_8$, although our results for helium reionization parameters do not significantly depend on this prior.

\begin{table}[b!]
  \vspace{-0.7cm}
  \begin{center}
    \caption{The detection signal-to-noise (SNR) of the (reconstructed) velocity and galaxy-density cross-correlation $P_{\hat{v}g}(k)$. Velocities are reconstructed from the kSZ tomography using VRO and MegaMapper surveys, together CMB measurements from Simons Observatory (SO), CMB-S4 and CMB-HD.}
    \vspace{-0.0cm}
    \label{tab:detection-SNR}
    \begin{tabular}{| l | l l l l |} 
    \hline
   \hline
    kSZ SNR & $\,\!z\!=1.9$ & 2.6 & 3.45 & 4.45  \\ 
    \hline
    VRO$+$CMB-HD & 1087 & 879 & 351 & 51 \\
    VRO$+$CMB-S4 & 186 & 126 & 48 & 7 \\ 
    VRO$+$CMB-SO & 87 & 59 & 23 & 4 \\
    \hline
    MegaMapper$+$CMB-HD & 1629 & 1051 & 453 & 129 \\
    MegaMapper$+$CMB-S4 & 254 & 154 & 78 & 31 \\
   \hline
   \hline
   \end{tabular}
  \end{center}
  \vspace{-0.4cm}
\end{table}

Fig.~\ref{fig:constraints1} demonstrates the measurement accuracy of the radial-velocity amplitude for combinations of VRO and MegaMapper surveys with CMB-S4 and CMB-HD, respectively. The signal-to-noise (SNR) of the kSZ tomography at each redshift is shown on Table~\ref{tab:detection-SNR} for a wider selection of CMB and LSS experimental configurations. Here, we define the detection signal-to-noise (SNR) of helium reionization as the SNR on $\Delta x_{\rm re}^{\rm He}$ (or $Y_p$) after marginalising over all other parameters. We find VRO and CMB-S4 can potentially detect helium reionization at $\{4\sigma, 8\sigma, 13\sigma\}$ significance for models 1-3 respectively. For the futuristic MegaMapper and CMB-HD, the detection SNR can reach $\{39\sigma, 56\sigma, 87\sigma\}$.

The sensitivities on the parameters describing the helium reionization{--given our fiducial model labelled 1--}are shown in Fig.~\ref{fig:Fisher_main}. The blue (orange) contours correspond to combination of VRO and CMB-S4 (MegaMapper and CMB-HD). In both cases we assume no prior information on the optical-depth and galaxy biases.\footnote{Note that better sensitivity on optical-depth or the galaxy biases only marginally improve the sensitivity of these experiments to helium reionization parameters, since the degeneracy between them and the bias parameters at each redshift is broken by the distinct redshift dependence of the models we consider.} The inner-most lighter-coloured contours assume $0.005$ prior on the $Y_p$ parameter, which can be provided from helium emission line measurements~\citep{Aver:2015iza} as well as potentially the CMB~\citep{Planck:2018vyg,CMB-S4:2016ple}. We find assuming priors on $Y_p$ improves the measurement accuracy on the other helium reionization parameters, most notably for VRO and CMB-S4. We show our forecasted sensitivities ($1\sigma$ errors) on a table inside the caption of Fig.~\ref{fig:Fisher_main}. We find the combination of VRO and CMB-S4 can measure the time of helium reionization at a precision that would allow distinguishing between models 1 and 3 and put potentially informative lower limits on duration of helium reionization. With MegaMapper and CMB-HD, we find kSZ tomography can measure the redshift and the duration of reionization at much higher significance, potentially allowing distinguishing between similar models.

{Interestingly, as demonstrated in Fig.~\ref{fig:Fisher_main}, we find the combination of MegaMapper and CMB-HD may have a sensitivity to $Y_p$ comparable to the accuracy of CMB and helium emission-line measurements. In order to assess further, we perform CMB forecasts on $Y_p$ using \texttt{FisherLens}, a publicly available\footnote{\url{https://github.com/ctrendafilova/FisherLens}} forecasting software~\citep{Hotinli:2021umk}. We take experimental specifications matching CMB-HD and with cosmological model parameters including the six standard $\Lambda$CDM parameters in addition to $N_{\rm eff}$, and $Y_p$. We observe a CMB-HD-like experiment {including the both temperature and polarization information} can be expected to achieve $\sigma(Y_p)\simeq0.004$ sensitivity, and that adding $\sigma(Y_p)\simeq0.006$ measurement from kSZ tomography from helium reionization can improve the error on $Y_p$ by $\sim15\%$. We find that this improvement leads to a $\sim10\%$ reduction in $N_{\rm eff}$ error, due to the partial breaking of the degeneracy suffered between the two parameters, suggesting kSZ measurements of helium reionization can potentially improve our understanding of relativistic species. 

\begin{figure}[t!]
    \hspace*{-0.6cm}\includegraphics[width=1.0\columnwidth]{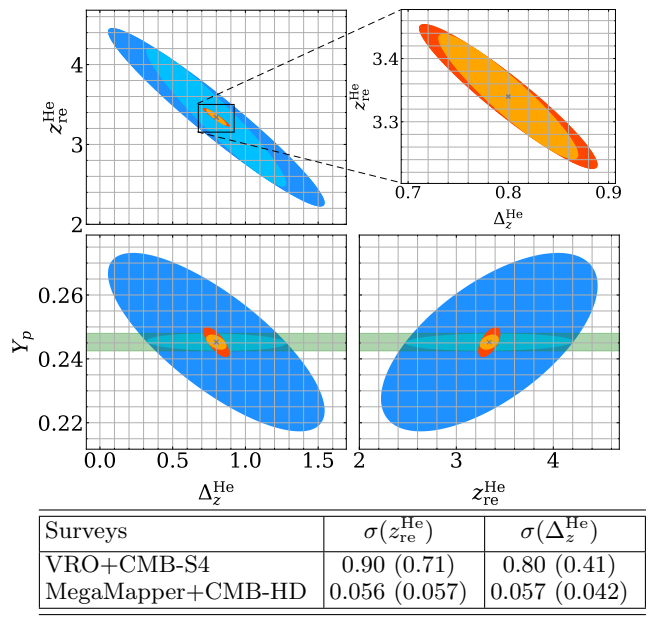}
    \vspace{-0.3cm}
    \begin{center}
    \begin{tabular}{ | l@{\hskip 6pt} | c@{\hskip 6pt} | c@{\hskip 6pt}  | } 
        \toprule
         Surveys & $\sigma(z_{\rm re}^{\rm He})$ & $\sigma(\Delta_{z}^{\rm He})$ \\ [0.5ex] 
     \hline
     VRO$+$CMB-S4 & 0.90 (0.71) & 0.80 (0.41)  \\
     MegaMapper$+$CMB-HD & 0.056 (0.057) & 0.057 (0.042) \\   
    \hline\hline
    \end{tabular}
    \end{center}
    \vspace*{-0.4cm}
    \caption{The sensitivities (1$\sigma$ errors) on the helium reionization parameters from two survey combinations: VRO and CMB-S4 shown with blue contours, and MegaMapper and CMB-HD, shown with orange contours. The inner lighter-coloured contours for each experiment corresponds to assuming $0.005$ prior on $Y_p$. The {table} shows the $1\sigma$ errors on $z_{\rm re}^{\rm He}$ and $\Delta_z^{\rm He}$
    The $1\sigma$ errors on $Y_p$ (without the $Y_p$ prior) are 0.06 and 0.006, respectively. The bracketed sensitivities correspond to assuming the $0.005$ prior on the helium fraction $Y_p$.}
    \vspace*{-0.5cm}
    \label{fig:Fisher_main} 
\end{figure}

We have omitted the potential effect of helium reionization on the selection function of high-$z$ quasars and galaxies. Across helium reionization, the ionizing processes can modulate the ultra-violet background fluctuations, the star formation and the absorption lines used for inferring the redshift with spectroscopic imaging surveys such as DESI and MegaMapper. Such effects can potentially cause significant changes that need to be taken into account in the selection function of these surveys, and likely need to be modelled for an unambiguous characterisation of helium reionization, as well as using more accurate inputs (such as the galaxy bias and number density) when performing forecasts in the future.

Throughout this paper, we used the so-called `box' formalism introduced in Ref.~\citep{Smith:2018bpn}. The benefit of this formalism is its simplicity; while using redshift bins on the light cone is likely a more accurate representation of kSZ tomography in practice, as discussed in Refs.~\citep{Cayuso:2021ljq,Deutsch:2017cja,Deutsch:2017ybc}, for example.\footnote{In addition, formalising kSZ tomography on the light cone has the benefit of better capturing the redshift dependence of the signal, particularly in the case of spectroscopic surveys where one can separate the redshift range in many bins, as well as better capturing the degrading effect of CMB foregrounds, for example. In our upcoming studies on this topic, we will focus on the light-cone formalism.} Here, our goal was to produce easy-to-reproduce forecasts that access and highlight the prospects of detecting and characterising helium {reionization}.

The epoch of helium reionization carries a large amount of information about astrophysics and cosmology that can potentially be accessed in the foreseeable future. As it occurs at lower redshifts, it allows the utilisation of the significant statistical power afforded by the LSS and CMB cross-correlation program-- a quality likely not shared with hydrogen reionization. 

Reconstructing velocities at high SNR with future surveys will provide precise tests of fundamental physics. We have shown here that this also provides a new path to detecting and characterising helium reionization.  These measurements will not require new experiments other than those being built or proposed, offering new opportunities and avenues for exploration for both cosmology and astrophysics. 

\section{Acknowledgements}

We thank Joel Meyers, Colin Hill, Alex van Engelen and Patrick Hall for useful discussions. This work was started in part at the Aspen Center for Physics, which is supported by National Science Foundation grant PHY-1607611. SCH is supported by the Horizon Fellowship from Johns Hopkins University. SCH also acknowledges the support of a grant from the Simons Foundation at the Aspen Center for Physics. SF is supported by the Physics Division of Lawrence Berkeley National Laboratory. GPH is supported by Brand \& Monica Fortner and the Canadian Insitute for Advanced Research. MCJ is supported by the National Science and Engineering Research Council through a Discovery grant. This research was supported in part by Perimeter Institute for Theoretical Physics. Research at Perimeter Institute is supported by the Government of Canada through the Department of Innovation, Science and Economic Development Canada and by the Province of Ontario through the Ministry of Research, Innovation and Science.  MK was supported by NSF Grant No.\ 2112699 and the Simons Foundation.

\bibliography{he_reion}

%merlin.mbs apsrev4-1.bst 2010-07-25 4.21a (PWD, AO, DPC) hacked
%Control: key (0)
%Control: author (72) initials jnrlst
%Control: editor formatted (1) identically to author
%Control: production of article title (-1) disabled
%Control: page (0) single
%Control: year (1) truncated
%Control: production of eprint (0) enabled
\begin{thebibliography}{57}%
\makeatletter
\providecommand \@ifxundefined [1]{%
 \@ifx{#1\undefined}
}%
\providecommand \@ifnum [1]{%
 \ifnum #1\expandafter \@firstoftwo
 \else \expandafter \@secondoftwo
 \fi
}%
\providecommand \@ifx [1]{%
 \ifx #1\expandafter \@firstoftwo
 \else \expandafter \@secondoftwo
 \fi
}%
\providecommand \natexlab [1]{#1}%
\providecommand \enquote  [1]{``#1''}%
\providecommand \bibnamefont  [1]{#1}%
\providecommand \bibfnamefont [1]{#1}%
\providecommand \citenamefont [1]{#1}%
\providecommand \href@noop [0]{\@secondoftwo}%
\providecommand \href [0]{\begingroup \@sanitize@url \@href}%
\providecommand \@href[1]{\@@startlink{#1}\@@href}%
\providecommand \@@href[1]{\endgroup#1\@@endlink}%
\providecommand \@sanitize@url [0]{\catcode `\\12\catcode `\$12\catcode
  `\&12\catcode `\#12\catcode `\^12\catcode `\_12\catcode `\%12\relax}%
\providecommand \@@startlink[1]{}%
\providecommand \@@endlink[0]{}%
\providecommand \url  [0]{\begingroup\@sanitize@url \@url }%
\providecommand \@url [1]{\endgroup\@href {#1}{\urlprefix }}%
\providecommand \urlprefix  [0]{URL }%
\providecommand \Eprint [0]{\href }%
\providecommand \doibase [0]{http://dx.doi.org/}%
\providecommand \selectlanguage [0]{\@gobble}%
\providecommand \bibinfo  [0]{\@secondoftwo}%
\providecommand \bibfield  [0]{\@secondoftwo}%
\providecommand \translation [1]{[#1]}%
\providecommand \BibitemOpen [0]{}%
\providecommand \bibitemStop [0]{}%
\providecommand \bibitemNoStop [0]{.\EOS\space}%
\providecommand \EOS [0]{\spacefactor3000\relax}%
\providecommand \BibitemShut  [1]{\csname bibitem#1\endcsname}%
\let\auto@bib@innerbib\@empty
%</preamble>
\bibitem [{\citenamefont {{Ross}}\ \emph {et~al.}(2013)\citenamefont {{Ross}},
  \citenamefont {{McGreer}}, \citenamefont {{White}}, \citenamefont
  {{Richards}}, \citenamefont {{Myers}}, \citenamefont
  {{Palanque-Delabrouille}}, \citenamefont {{Strauss}}, \citenamefont
  {{Anderson}}, \citenamefont {{Shen}}, \citenamefont {{Brandt}}, \citenamefont
  {{Y{\`e}che}}, \citenamefont {{Swanson}} \emph {et~al.}}]{Ross2013}%
  \BibitemOpen
  \bibfield  {author} {\bibinfo {author} {\bibfnamefont {N.~P.}\ \bibnamefont
  {{Ross}}}, \bibinfo {author} {\bibfnamefont {I.~D.}\ \bibnamefont
  {{McGreer}}}, \bibinfo {author} {\bibfnamefont {M.}~\bibnamefont {{White}}},
  \bibinfo {author} {\bibfnamefont {G.~T.}\ \bibnamefont {{Richards}}},
  \bibinfo {author} {\bibfnamefont {A.~D.}\ \bibnamefont {{Myers}}}, \bibinfo
  {author} {\bibfnamefont {N.}~\bibnamefont {{Palanque-Delabrouille}}},
  \bibinfo {author} {\bibfnamefont {M.~A.}\ \bibnamefont {{Strauss}}}, \bibinfo
  {author} {\bibfnamefont {S.~F.}\ \bibnamefont {{Anderson}}}, \bibinfo
  {author} {\bibfnamefont {Y.}~\bibnamefont {{Shen}}}, \bibinfo {author}
  {\bibfnamefont {W.~N.}\ \bibnamefont {{Brandt}}}, \bibinfo {author}
  {\bibfnamefont {C.}~\bibnamefont {{Y{\`e}che}}}, \bibinfo {author}
  {\bibfnamefont {M.~E.~C.}\ \bibnamefont {{Swanson}}},  \emph {et~al.},\
  }\href {\doibase 10.1088/0004-637X/773/1/14} {\bibfield  {journal} {\bibinfo
  {journal} {\apj}\ }\textbf {\bibinfo {volume} {773}},\ \bibinfo {eid} {14}
  (\bibinfo {year} {2013})},\ \Eprint {http://arxiv.org/abs/1210.6389}
  {arXiv:1210.6389 [astro-ph.CO]} \BibitemShut {NoStop}%
\bibitem [{\citenamefont {{Masters}}\ \emph {et~al.}(2012)\citenamefont
  {{Masters}}, \citenamefont {{Capak}}, \citenamefont {{Salvato}},
  \citenamefont {{Civano}}, \citenamefont {{Mobasher}}, \citenamefont
  {{Siana}}, \citenamefont {{Hasinger}}, \citenamefont {{Impey}}, \citenamefont
  {{Nagao}}, \citenamefont {{Trump}}, \citenamefont {{Ikeda}}, \citenamefont
  {{Elvis}},\ and\ \citenamefont {{Scoville}}}]{Masters2012}%
  \BibitemOpen
  \bibfield  {author} {\bibinfo {author} {\bibfnamefont {D.}~\bibnamefont
  {{Masters}}}, \bibinfo {author} {\bibfnamefont {P.}~\bibnamefont {{Capak}}},
  \bibinfo {author} {\bibfnamefont {M.}~\bibnamefont {{Salvato}}}, \bibinfo
  {author} {\bibfnamefont {F.}~\bibnamefont {{Civano}}}, \bibinfo {author}
  {\bibfnamefont {B.}~\bibnamefont {{Mobasher}}}, \bibinfo {author}
  {\bibfnamefont {B.}~\bibnamefont {{Siana}}}, \bibinfo {author} {\bibfnamefont
  {G.}~\bibnamefont {{Hasinger}}}, \bibinfo {author} {\bibfnamefont {C.~D.}\
  \bibnamefont {{Impey}}}, \bibinfo {author} {\bibfnamefont {T.}~\bibnamefont
  {{Nagao}}}, \bibinfo {author} {\bibfnamefont {J.~R.}\ \bibnamefont
  {{Trump}}}, \bibinfo {author} {\bibfnamefont {H.}~\bibnamefont {{Ikeda}}},
  \bibinfo {author} {\bibfnamefont {M.}~\bibnamefont {{Elvis}}}, \ and\
  \bibinfo {author} {\bibfnamefont {N.}~\bibnamefont {{Scoville}}},\ }\href
  {\doibase 10.1088/0004-637X/755/2/169} {\bibfield  {journal} {\bibinfo
  {journal} {\apj}\ }\textbf {\bibinfo {volume} {755}},\ \bibinfo {eid} {169}
  (\bibinfo {year} {2012})},\ \Eprint {http://arxiv.org/abs/1207.2154}
  {arXiv:1207.2154 [astro-ph.CO]} \BibitemShut {NoStop}%
\bibitem [{\citenamefont {{McGreer}}\ \emph {et~al.}(2013)\citenamefont
  {{McGreer}}, \citenamefont {{Jiang}}, \citenamefont {{Fan}}, \citenamefont
  {{Richards}}, \citenamefont {{Strauss}}, \citenamefont {{Ross}},
  \citenamefont {{White}}, \citenamefont {{Shen}}, \citenamefont {{Schneider}},
  \citenamefont {{Myers}}, \citenamefont {{Brandt}}, \citenamefont {{DeGraf}},
  \citenamefont {{Glikman}}, \citenamefont {{Ge}},\ and\ \citenamefont
  {{Streblyanska}}}]{McGreer2013}%
  \BibitemOpen
  \bibfield  {author} {\bibinfo {author} {\bibfnamefont {I.~D.}\ \bibnamefont
  {{McGreer}}}, \bibinfo {author} {\bibfnamefont {L.}~\bibnamefont {{Jiang}}},
  \bibinfo {author} {\bibfnamefont {X.}~\bibnamefont {{Fan}}}, \bibinfo
  {author} {\bibfnamefont {G.~T.}\ \bibnamefont {{Richards}}}, \bibinfo
  {author} {\bibfnamefont {M.~A.}\ \bibnamefont {{Strauss}}}, \bibinfo {author}
  {\bibfnamefont {N.~P.}\ \bibnamefont {{Ross}}}, \bibinfo {author}
  {\bibfnamefont {M.}~\bibnamefont {{White}}}, \bibinfo {author} {\bibfnamefont
  {Y.}~\bibnamefont {{Shen}}}, \bibinfo {author} {\bibfnamefont {D.~P.}\
  \bibnamefont {{Schneider}}}, \bibinfo {author} {\bibfnamefont {A.~D.}\
  \bibnamefont {{Myers}}}, \bibinfo {author} {\bibfnamefont {W.~N.}\
  \bibnamefont {{Brandt}}}, \bibinfo {author} {\bibfnamefont {C.}~\bibnamefont
  {{DeGraf}}}, \bibinfo {author} {\bibfnamefont {E.}~\bibnamefont {{Glikman}}},
  \bibinfo {author} {\bibfnamefont {J.}~\bibnamefont {{Ge}}}, \ and\ \bibinfo
  {author} {\bibfnamefont {A.}~\bibnamefont {{Streblyanska}}},\ }\href
  {\doibase 10.1088/0004-637X/768/2/105} {\bibfield  {journal} {\bibinfo
  {journal} {\apj}\ }\textbf {\bibinfo {volume} {768}},\ \bibinfo {eid} {105}
  (\bibinfo {year} {2013})},\ \Eprint {http://arxiv.org/abs/1212.4493}
  {arXiv:1212.4493 [astro-ph.CO]} \BibitemShut {NoStop}%
\bibitem [{\citenamefont {{McGreer}}\ \emph {et~al.}(2018)\citenamefont
  {{McGreer}}, \citenamefont {{Fan}}, \citenamefont {{Jiang}},\ and\
  \citenamefont {{Cai}}}]{McGreer2018}%
  \BibitemOpen
  \bibfield  {author} {\bibinfo {author} {\bibfnamefont {I.~D.}\ \bibnamefont
  {{McGreer}}}, \bibinfo {author} {\bibfnamefont {X.}~\bibnamefont {{Fan}}},
  \bibinfo {author} {\bibfnamefont {L.}~\bibnamefont {{Jiang}}}, \ and\
  \bibinfo {author} {\bibfnamefont {Z.}~\bibnamefont {{Cai}}},\ }\href
  {\doibase 10.3847/1538-3881/aaaab4} {\bibfield  {journal} {\bibinfo
  {journal} {Astron. J.}\ }\textbf {\bibinfo {volume} {155}},\ \bibinfo {eid}
  {131} (\bibinfo {year} {2018})},\ \Eprint {http://arxiv.org/abs/1710.09390}
  {arXiv:1710.09390 [astro-ph.GA]} \BibitemShut {NoStop}%
\bibitem [{\citenamefont {{Pan}}\ \emph {et~al.}(2022)\citenamefont {{Pan}},
  \citenamefont {{Jiang}}, \citenamefont {{Fan}}, \citenamefont {{Wu}},\ and\
  \citenamefont {{Yang}}}]{Pan2022}%
  \BibitemOpen
  \bibfield  {author} {\bibinfo {author} {\bibfnamefont {Z.}~\bibnamefont
  {{Pan}}}, \bibinfo {author} {\bibfnamefont {L.}~\bibnamefont {{Jiang}}},
  \bibinfo {author} {\bibfnamefont {X.}~\bibnamefont {{Fan}}}, \bibinfo
  {author} {\bibfnamefont {J.}~\bibnamefont {{Wu}}}, \ and\ \bibinfo {author}
  {\bibfnamefont {J.}~\bibnamefont {{Yang}}},\ }\href {\doibase
  10.3847/1538-4357/ac5aab} {\bibfield  {journal} {\bibinfo  {journal} {\apj}\
  }\textbf {\bibinfo {volume} {928}},\ \bibinfo {eid} {172} (\bibinfo {year}
  {2022})},\ \Eprint {http://arxiv.org/abs/2112.07801} {arXiv:2112.07801
  [astro-ph.GA]} \BibitemShut {NoStop}%
\bibitem [{\citenamefont {Shen}\ and\ \citenamefont {Ho}(2014)}]{Shen:2014rka}%
  \BibitemOpen
  \bibfield  {author} {\bibinfo {author} {\bibfnamefont {Y.}~\bibnamefont
  {Shen}}\ and\ \bibinfo {author} {\bibfnamefont {L.~C.}\ \bibnamefont {Ho}},\
  }\href {\doibase 10.1038/nature13712} {\bibfield  {journal} {\bibinfo
  {journal} {Nature}\ }\textbf {\bibinfo {volume} {513}},\ \bibinfo {pages}
  {210} (\bibinfo {year} {2014})},\ \Eprint {http://arxiv.org/abs/1409.2887}
  {arXiv:1409.2887 [astro-ph.GA]} \BibitemShut {NoStop}%
\bibitem [{\citenamefont {Hopkins}\ \emph {et~al.}(2007)\citenamefont
  {Hopkins}, \citenamefont {Lidz}, \citenamefont {Hernquist}, \citenamefont
  {Coil}, \citenamefont {Myers}, \citenamefont {Cox},\ and\ \citenamefont
  {Spergel}}]{Hopkins:2006vv}%
  \BibitemOpen
  \bibfield  {author} {\bibinfo {author} {\bibfnamefont {P.~F.}\ \bibnamefont
  {Hopkins}}, \bibinfo {author} {\bibfnamefont {A.}~\bibnamefont {Lidz}},
  \bibinfo {author} {\bibfnamefont {L.}~\bibnamefont {Hernquist}}, \bibinfo
  {author} {\bibfnamefont {A.~L.}\ \bibnamefont {Coil}}, \bibinfo {author}
  {\bibfnamefont {A.~D.}\ \bibnamefont {Myers}}, \bibinfo {author}
  {\bibfnamefont {T.~J.}\ \bibnamefont {Cox}}, \ and\ \bibinfo {author}
  {\bibfnamefont {D.~N.}\ \bibnamefont {Spergel}},\ }\href {\doibase
  10.1086/517512} {\bibfield  {journal} {\bibinfo  {journal} {Astrophys. J.}\
  }\textbf {\bibinfo {volume} {662}},\ \bibinfo {pages} {110} (\bibinfo {year}
  {2007})},\ \Eprint {http://arxiv.org/abs/astro-ph/0611792}
  {arXiv:astro-ph/0611792} \BibitemShut {NoStop}%
\bibitem [{\citenamefont {{Schmidt}}\ \emph {et~al.}(2017)\citenamefont
  {{Schmidt}}, \citenamefont {{Worseck}}, \citenamefont {{Hennawi}},
  \citenamefont {{Prochaska}},\ and\ \citenamefont {{Crighton}}}]{Schmidt2017}%
  \BibitemOpen
  \bibfield  {author} {\bibinfo {author} {\bibfnamefont {T.~M.}\ \bibnamefont
  {{Schmidt}}}, \bibinfo {author} {\bibfnamefont {G.}~\bibnamefont
  {{Worseck}}}, \bibinfo {author} {\bibfnamefont {J.~F.}\ \bibnamefont
  {{Hennawi}}}, \bibinfo {author} {\bibfnamefont {J.~X.}\ \bibnamefont
  {{Prochaska}}}, \ and\ \bibinfo {author} {\bibfnamefont {N.~H.~M.}\
  \bibnamefont {{Crighton}}},\ }\href {\doibase 10.3847/1538-4357/aa83ac}
  {\bibfield  {journal} {\bibinfo  {journal} {\apj}\ }\textbf {\bibinfo
  {volume} {847}},\ \bibinfo {eid} {81} (\bibinfo {year} {2017})},\ \Eprint
  {http://arxiv.org/abs/1701.08769} {arXiv:1701.08769 [astro-ph.GA]}
  \BibitemShut {NoStop}%
\bibitem [{\citenamefont {Inayoshi}\ \emph {et~al.}(2020)\citenamefont
  {Inayoshi}, \citenamefont {Visbal},\ and\ \citenamefont
  {Haiman}}]{Inayoshi:2019fun}%
  \BibitemOpen
  \bibfield  {author} {\bibinfo {author} {\bibfnamefont {K.}~\bibnamefont
  {Inayoshi}}, \bibinfo {author} {\bibfnamefont {E.}~\bibnamefont {Visbal}}, \
  and\ \bibinfo {author} {\bibfnamefont {Z.}~\bibnamefont {Haiman}},\ }\href
  {\doibase 10.1146/annurev-astro-120419-014455} {\bibfield  {journal}
  {\bibinfo  {journal} {Ann. Rev. Astron. Astrophys.}\ }\textbf {\bibinfo
  {volume} {58}},\ \bibinfo {pages} {27} (\bibinfo {year} {2020})},\ \Eprint
  {http://arxiv.org/abs/1911.05791} {arXiv:1911.05791 [astro-ph.GA]}
  \BibitemShut {NoStop}%
\bibitem [{\citenamefont {{Richards}}\ \emph {et~al.}(2006)\citenamefont
  {{Richards}}, \citenamefont {{Strauss}}, \citenamefont {{Fan}}, \citenamefont
  {{Hall}}, \citenamefont {{Jester}}, \citenamefont {{Schneider}},
  \citenamefont {{Vanden Berk}}, \citenamefont {{Stoughton}} \emph
  {et~al.}}]{Richards2006}%
  \BibitemOpen
  \bibfield  {author} {\bibinfo {author} {\bibfnamefont {G.~T.}\ \bibnamefont
  {{Richards}}}, \bibinfo {author} {\bibfnamefont {M.~A.}\ \bibnamefont
  {{Strauss}}}, \bibinfo {author} {\bibfnamefont {X.}~\bibnamefont {{Fan}}},
  \bibinfo {author} {\bibfnamefont {P.~B.}\ \bibnamefont {{Hall}}}, \bibinfo
  {author} {\bibfnamefont {S.}~\bibnamefont {{Jester}}}, \bibinfo {author}
  {\bibfnamefont {D.~P.}\ \bibnamefont {{Schneider}}}, \bibinfo {author}
  {\bibfnamefont {D.~E.}\ \bibnamefont {{Vanden Berk}}}, \bibinfo {author}
  {\bibfnamefont {C.}~\bibnamefont {{Stoughton}}},  \emph {et~al.},\ }\href
  {\doibase 10.1086/503559} {\bibfield  {journal} {\bibinfo  {journal} {Astron.
  J.}\ }\textbf {\bibinfo {volume} {131}},\ \bibinfo {pages} {2766} (\bibinfo
  {year} {2006})},\ \Eprint {http://arxiv.org/abs/astro-ph/0601434}
  {arXiv:astro-ph/0601434 [astro-ph]} \BibitemShut {NoStop}%
\bibitem [{\citenamefont {{Jakobsen}}\ \emph {et~al.}(1994)\citenamefont
  {{Jakobsen}}, \citenamefont {{Boksenberg}}, \citenamefont {{Deharveng}},
  \citenamefont {{Greenfield}}, \citenamefont {{Jedrzejewski}},\ and\
  \citenamefont {{Paresce}}}]{Jakobsen1994}%
  \BibitemOpen
  \bibfield  {author} {\bibinfo {author} {\bibfnamefont {P.}~\bibnamefont
  {{Jakobsen}}}, \bibinfo {author} {\bibfnamefont {A.}~\bibnamefont
  {{Boksenberg}}}, \bibinfo {author} {\bibfnamefont {J.~M.}\ \bibnamefont
  {{Deharveng}}}, \bibinfo {author} {\bibfnamefont {P.}~\bibnamefont
  {{Greenfield}}}, \bibinfo {author} {\bibfnamefont {R.}~\bibnamefont
  {{Jedrzejewski}}}, \ and\ \bibinfo {author} {\bibfnamefont {F.}~\bibnamefont
  {{Paresce}}},\ }\href {\doibase 10.1038/370035a0} {\bibfield  {journal}
  {\bibinfo  {journal} {\nat}\ }\textbf {\bibinfo {volume} {370}},\ \bibinfo
  {pages} {35} (\bibinfo {year} {1994})}\BibitemShut {NoStop}%
\bibitem [{\citenamefont {{Zheng}}\ \emph {et~al.}(2008)\citenamefont
  {{Zheng}}, \citenamefont {{Meiksin}}, \citenamefont {{Pifko}}, \citenamefont
  {{Anderson}}, \citenamefont {{Hogan}}, \citenamefont {{Tittley}},
  \citenamefont {{Kriss}}, \citenamefont {{Chiu}}, \citenamefont {{Schneider}},
  \citenamefont {{York}},\ and\ \citenamefont {{Weinberg}}}]{Zheng2008}%
  \BibitemOpen
  \bibfield  {author} {\bibinfo {author} {\bibfnamefont {W.}~\bibnamefont
  {{Zheng}}}, \bibinfo {author} {\bibfnamefont {A.}~\bibnamefont {{Meiksin}}},
  \bibinfo {author} {\bibfnamefont {K.}~\bibnamefont {{Pifko}}}, \bibinfo
  {author} {\bibfnamefont {S.~F.}\ \bibnamefont {{Anderson}}}, \bibinfo
  {author} {\bibfnamefont {C.~J.}\ \bibnamefont {{Hogan}}}, \bibinfo {author}
  {\bibfnamefont {E.}~\bibnamefont {{Tittley}}}, \bibinfo {author}
  {\bibfnamefont {G.~A.}\ \bibnamefont {{Kriss}}}, \bibinfo {author}
  {\bibfnamefont {K.}~\bibnamefont {{Chiu}}}, \bibinfo {author} {\bibfnamefont
  {D.~P.}\ \bibnamefont {{Schneider}}}, \bibinfo {author} {\bibfnamefont
  {D.~G.}\ \bibnamefont {{York}}}, \ and\ \bibinfo {author} {\bibfnamefont
  {D.~H.}\ \bibnamefont {{Weinberg}}},\ }\href {\doibase 10.1086/590384}
  {\bibfield  {journal} {\bibinfo  {journal} {\apj}\ }\textbf {\bibinfo
  {volume} {686}},\ \bibinfo {pages} {195} (\bibinfo {year}
  {2008})}\BibitemShut {NoStop}%
\bibitem [{\citenamefont {{Syphers}}\ and\ \citenamefont
  {{Shull}}(2014)}]{Syphers2014}%
  \BibitemOpen
  \bibfield  {author} {\bibinfo {author} {\bibfnamefont {D.}~\bibnamefont
  {{Syphers}}}\ and\ \bibinfo {author} {\bibfnamefont {J.~M.}\ \bibnamefont
  {{Shull}}},\ }\href {\doibase 10.1088/0004-637X/784/1/42} {\bibfield
  {journal} {\bibinfo  {journal} {\apj}\ }\textbf {\bibinfo {volume} {784}},\
  \bibinfo {eid} {42} (\bibinfo {year} {2014})}\BibitemShut {NoStop}%
\bibitem [{\citenamefont {{Calura}}\ \emph {et~al.}(2012)\citenamefont
  {{Calura}}, \citenamefont {{Tescari}}, \citenamefont {{D'Odorico}},
  \citenamefont {{Viel}}, \citenamefont {{Cristiani}}, \citenamefont {{Kim}},\
  and\ \citenamefont {{Bolton}}}]{Calura2012}%
  \BibitemOpen
  \bibfield  {author} {\bibinfo {author} {\bibfnamefont {F.}~\bibnamefont
  {{Calura}}}, \bibinfo {author} {\bibfnamefont {E.}~\bibnamefont {{Tescari}}},
  \bibinfo {author} {\bibfnamefont {V.}~\bibnamefont {{D'Odorico}}}, \bibinfo
  {author} {\bibfnamefont {M.}~\bibnamefont {{Viel}}}, \bibinfo {author}
  {\bibfnamefont {S.}~\bibnamefont {{Cristiani}}}, \bibinfo {author}
  {\bibfnamefont {T.~S.}\ \bibnamefont {{Kim}}}, \ and\ \bibinfo {author}
  {\bibfnamefont {J.~S.}\ \bibnamefont {{Bolton}}},\ }\href {\doibase
  10.1111/j.1365-2966.2012.20811.x} {\bibfield  {journal} {\bibinfo  {journal}
  {MNRAS}\ }\textbf {\bibinfo {volume} {422}},\ \bibinfo {pages} {3019}
  (\bibinfo {year} {2012})},\ \Eprint {http://arxiv.org/abs/1201.5121}
  {arXiv:1201.5121 [astro-ph.CO]} \BibitemShut {NoStop}%
\bibitem [{\citenamefont {{Viel}}\ \emph {et~al.}(2013)\citenamefont {{Viel}},
  \citenamefont {{Becker}}, \citenamefont {{Bolton}},\ and\ \citenamefont
  {{Haehnelt}}}]{Viel2013}%
  \BibitemOpen
  \bibfield  {author} {\bibinfo {author} {\bibfnamefont {M.}~\bibnamefont
  {{Viel}}}, \bibinfo {author} {\bibfnamefont {G.~D.}\ \bibnamefont
  {{Becker}}}, \bibinfo {author} {\bibfnamefont {J.~S.}\ \bibnamefont
  {{Bolton}}}, \ and\ \bibinfo {author} {\bibfnamefont {M.~G.}\ \bibnamefont
  {{Haehnelt}}},\ }\href {\doibase 10.1103/PhysRevD.88.043502} {\bibfield
  {journal} {\bibinfo  {journal} {\prd}\ }\textbf {\bibinfo {volume} {88}},\
  \bibinfo {eid} {043502} (\bibinfo {year} {2013})},\ \Eprint
  {http://arxiv.org/abs/1306.2314} {arXiv:1306.2314 [astro-ph.CO]} \BibitemShut
  {NoStop}%
\bibitem [{\citenamefont {{Boera}}\ \emph {et~al.}(2016)\citenamefont
  {{Boera}}, \citenamefont {{Murphy}}, \citenamefont {{Becker}},\ and\
  \citenamefont {{Bolton}}}]{Boera2016}%
  \BibitemOpen
  \bibfield  {author} {\bibinfo {author} {\bibfnamefont {E.}~\bibnamefont
  {{Boera}}}, \bibinfo {author} {\bibfnamefont {M.~T.}\ \bibnamefont
  {{Murphy}}}, \bibinfo {author} {\bibfnamefont {G.~D.}\ \bibnamefont
  {{Becker}}}, \ and\ \bibinfo {author} {\bibfnamefont {J.~S.}\ \bibnamefont
  {{Bolton}}},\ }\href {\doibase 10.1093/mnrasl/slv172} {\bibfield  {journal}
  {\bibinfo  {journal} {MNRAS}\ }\textbf {\bibinfo {volume} {456}},\ \bibinfo
  {pages} {L79} (\bibinfo {year} {2016})},\ \Eprint
  {http://arxiv.org/abs/1510.08857} {arXiv:1510.08857 [astro-ph.CO]}
  \BibitemShut {NoStop}%
\bibitem [{\citenamefont {{Upton Sanderbeck}}\ \emph
  {et~al.}(2016)\citenamefont {{Upton Sanderbeck}}, \citenamefont
  {{D'Aloisio}},\ and\ \citenamefont {{McQuinn}}}]{UptonSanderbeck2016}%
  \BibitemOpen
  \bibfield  {author} {\bibinfo {author} {\bibfnamefont {P.~R.}\ \bibnamefont
  {{Upton Sanderbeck}}}, \bibinfo {author} {\bibfnamefont {A.}~\bibnamefont
  {{D'Aloisio}}}, \ and\ \bibinfo {author} {\bibfnamefont {M.~J.}\ \bibnamefont
  {{McQuinn}}},\ }\href {\doibase 10.1093/mnras/stw1117} {\bibfield  {journal}
  {\bibinfo  {journal} {MNRAS}\ }\textbf {\bibinfo {volume} {460}},\ \bibinfo
  {pages} {1885} (\bibinfo {year} {2016})},\ \Eprint
  {http://arxiv.org/abs/1511.05992} {arXiv:1511.05992 [astro-ph.CO]}
  \BibitemShut {NoStop}%
\bibitem [{\citenamefont {{La Plante}}\ \emph {et~al.}(2017)\citenamefont {{La
  Plante}}, \citenamefont {{Trac}}, \citenamefont {{Croft}},\ and\
  \citenamefont {{Cen}}}]{LaPlante2017}%
  \BibitemOpen
  \bibfield  {author} {\bibinfo {author} {\bibfnamefont {P.}~\bibnamefont {{La
  Plante}}}, \bibinfo {author} {\bibfnamefont {H.}~\bibnamefont {{Trac}}},
  \bibinfo {author} {\bibfnamefont {R.}~\bibnamefont {{Croft}}}, \ and\
  \bibinfo {author} {\bibfnamefont {R.}~\bibnamefont {{Cen}}},\ }\href
  {\doibase 10.3847/1538-4357/aa7136} {\bibfield  {journal} {\bibinfo
  {journal} {\apj}\ }\textbf {\bibinfo {volume} {841}},\ \bibinfo {eid} {87}
  (\bibinfo {year} {2017})},\ \Eprint {http://arxiv.org/abs/1610.02047}
  {arXiv:1610.02047 [astro-ph.CO]} \BibitemShut {NoStop}%
\bibitem [{\citenamefont {{La Plante}}\ \emph {et~al.}(2018)\citenamefont {{La
  Plante}}, \citenamefont {{Trac}}, \citenamefont {{Croft}},\ and\
  \citenamefont {{Cen}}}]{LaPlante2018}%
  \BibitemOpen
  \bibfield  {author} {\bibinfo {author} {\bibfnamefont {P.}~\bibnamefont {{La
  Plante}}}, \bibinfo {author} {\bibfnamefont {H.}~\bibnamefont {{Trac}}},
  \bibinfo {author} {\bibfnamefont {R.}~\bibnamefont {{Croft}}}, \ and\
  \bibinfo {author} {\bibfnamefont {R.}~\bibnamefont {{Cen}}},\ }\href
  {\doibase 10.3847/1538-4357/aae693} {\bibfield  {journal} {\bibinfo
  {journal} {\apj}\ }\textbf {\bibinfo {volume} {868}},\ \bibinfo {eid} {106}
  (\bibinfo {year} {2018})},\ \Eprint {http://arxiv.org/abs/1710.03286}
  {arXiv:1710.03286 [astro-ph.CO]} \BibitemShut {NoStop}%
\bibitem [{\citenamefont {{Bolton}}\ \emph {et~al.}(2017)\citenamefont
  {{Bolton}}, \citenamefont {{Puchwein}}, \citenamefont {{Sijacki}},
  \citenamefont {{Haehnelt}}, \citenamefont {{Kim}}, \citenamefont {{Meiksin}},
  \citenamefont {{Regan}},\ and\ \citenamefont {{Viel}}}]{Bolton2017}%
  \BibitemOpen
  \bibfield  {author} {\bibinfo {author} {\bibfnamefont {J.~S.}\ \bibnamefont
  {{Bolton}}}, \bibinfo {author} {\bibfnamefont {E.}~\bibnamefont
  {{Puchwein}}}, \bibinfo {author} {\bibfnamefont {D.}~\bibnamefont
  {{Sijacki}}}, \bibinfo {author} {\bibfnamefont {M.~G.}\ \bibnamefont
  {{Haehnelt}}}, \bibinfo {author} {\bibfnamefont {T.-S.}\ \bibnamefont
  {{Kim}}}, \bibinfo {author} {\bibfnamefont {A.}~\bibnamefont {{Meiksin}}},
  \bibinfo {author} {\bibfnamefont {J.~A.}\ \bibnamefont {{Regan}}}, \ and\
  \bibinfo {author} {\bibfnamefont {M.}~\bibnamefont {{Viel}}},\ }\href
  {\doibase 10.1093/mnras/stw2397} {\bibfield  {journal} {\bibinfo  {journal}
  {MNRAS}\ }\textbf {\bibinfo {volume} {464}},\ \bibinfo {pages} {897}
  (\bibinfo {year} {2017})},\ \Eprint {http://arxiv.org/abs/1605.03462}
  {arXiv:1605.03462 [astro-ph.CO]} \BibitemShut {NoStop}%
\bibitem [{\citenamefont {{Syphers}}\ \emph {et~al.}(2012)\citenamefont
  {{Syphers}}, \citenamefont {{Anderson}}, \citenamefont {{Zheng}},
  \citenamefont {{Meiksin}}, \citenamefont {{Schneider}},\ and\ \citenamefont
  {{York}}}]{Syphers2012}%
  \BibitemOpen
  \bibfield  {author} {\bibinfo {author} {\bibfnamefont {D.}~\bibnamefont
  {{Syphers}}}, \bibinfo {author} {\bibfnamefont {S.~F.}\ \bibnamefont
  {{Anderson}}}, \bibinfo {author} {\bibfnamefont {W.}~\bibnamefont {{Zheng}}},
  \bibinfo {author} {\bibfnamefont {A.}~\bibnamefont {{Meiksin}}}, \bibinfo
  {author} {\bibfnamefont {D.~P.}\ \bibnamefont {{Schneider}}}, \ and\ \bibinfo
  {author} {\bibfnamefont {D.~G.}\ \bibnamefont {{York}}},\ }\href {\doibase
  10.1088/0004-6256/143/4/100} {\bibfield  {journal} {\bibinfo  {journal}
  {Astron. J.}\ }\textbf {\bibinfo {volume} {143}},\ \bibinfo {eid} {100}
  (\bibinfo {year} {2012})},\ \Eprint {http://arxiv.org/abs/1202.0236}
  {arXiv:1202.0236 [astro-ph.CO]} \BibitemShut {NoStop}%
\bibitem [{\citenamefont {{Linder}}(2020)}]{2020PhRvD.101j3019L}%
  \BibitemOpen
  \bibfield  {author} {\bibinfo {author} {\bibfnamefont {E.~V.}\ \bibnamefont
  {{Linder}}},\ }\href {\doibase 10.1103/PhysRevD.101.103019} {\bibfield
  {journal} {\bibinfo  {journal} {\prd}\ }\textbf {\bibinfo {volume} {101}},\
  \bibinfo {eid} {103019} (\bibinfo {year} {2020})},\ \Eprint
  {http://arxiv.org/abs/2001.11517} {arXiv:2001.11517 [astro-ph.CO]}
  \BibitemShut {NoStop}%
\bibitem [{\citenamefont {{Bhattacharya}}\ \emph {et~al.}(2021)\citenamefont
  {{Bhattacharya}}, \citenamefont {{Kumar}},\ and\ \citenamefont
  {{Linder}}}]{2021PhRvD.103j3526B}%
  \BibitemOpen
  \bibfield  {author} {\bibinfo {author} {\bibfnamefont {M.}~\bibnamefont
  {{Bhattacharya}}}, \bibinfo {author} {\bibfnamefont {P.}~\bibnamefont
  {{Kumar}}}, \ and\ \bibinfo {author} {\bibfnamefont {E.~V.}\ \bibnamefont
  {{Linder}}},\ }\href {\doibase 10.1103/PhysRevD.103.103526} {\bibfield
  {journal} {\bibinfo  {journal} {\prd}\ }\textbf {\bibinfo {volume} {103}},\
  \bibinfo {eid} {103526} (\bibinfo {year} {2021})},\ \Eprint
  {http://arxiv.org/abs/2010.14530} {arXiv:2010.14530 [astro-ph.CO]}
  \BibitemShut {NoStop}%
\bibitem [{\citenamefont {{Sunyaev}}\ and\ \citenamefont
  {{Zeldovich}}(1972)}]{Sunyaev1972}%
  \BibitemOpen
  \bibfield  {author} {\bibinfo {author} {\bibfnamefont {R.~A.}\ \bibnamefont
  {{Sunyaev}}}\ and\ \bibinfo {author} {\bibfnamefont {Y.~B.}\ \bibnamefont
  {{Zeldovich}}},\ }\href@noop {} {\bibfield  {journal} {\bibinfo  {journal}
  {Comments on Astrophysics and Space Physics}\ }\textbf {\bibinfo {volume}
  {4}},\ \bibinfo {pages} {173} (\bibinfo {year} {1972})}\BibitemShut {NoStop}%
\bibitem [{\citenamefont {Deutsch}\ \emph {et~al.}(2018)\citenamefont
  {Deutsch}, \citenamefont {Dimastrogiovanni}, \citenamefont {Johnson},
  \citenamefont {M\"unchmeyer},\ and\ \citenamefont
  {Terrana}}]{Deutsch:2017ybc}%
  \BibitemOpen
  \bibfield  {author} {\bibinfo {author} {\bibfnamefont {A.-S.}\ \bibnamefont
  {Deutsch}}, \bibinfo {author} {\bibfnamefont {E.}~\bibnamefont
  {Dimastrogiovanni}}, \bibinfo {author} {\bibfnamefont {M.~C.}\ \bibnamefont
  {Johnson}}, \bibinfo {author} {\bibfnamefont {M.}~\bibnamefont
  {M\"unchmeyer}}, \ and\ \bibinfo {author} {\bibfnamefont {A.}~\bibnamefont
  {Terrana}},\ }\href {\doibase 10.1103/PhysRevD.98.123501} {\bibfield
  {journal} {\bibinfo  {journal} {Phys. Rev. D}\ }\textbf {\bibinfo {volume}
  {98}},\ \bibinfo {pages} {123501} (\bibinfo {year} {2018})},\ \Eprint
  {http://arxiv.org/abs/1707.08129} {arXiv:1707.08129 [astro-ph.CO]}
  \BibitemShut {NoStop}%
\bibitem [{\citenamefont {Smith}\ \emph {et~al.}(2018)\citenamefont {Smith},
  \citenamefont {Madhavacheril}, \citenamefont {M\"unchmeyer}, \citenamefont
  {Ferraro}, \citenamefont {Giri},\ and\ \citenamefont
  {Johnson}}]{Smith:2018bpn}%
  \BibitemOpen
  \bibfield  {author} {\bibinfo {author} {\bibfnamefont {K.~M.}\ \bibnamefont
  {Smith}}, \bibinfo {author} {\bibfnamefont {M.~S.}\ \bibnamefont
  {Madhavacheril}}, \bibinfo {author} {\bibfnamefont {M.}~\bibnamefont
  {M\"unchmeyer}}, \bibinfo {author} {\bibfnamefont {S.}~\bibnamefont
  {Ferraro}}, \bibinfo {author} {\bibfnamefont {U.}~\bibnamefont {Giri}}, \
  and\ \bibinfo {author} {\bibfnamefont {M.~C.}\ \bibnamefont {Johnson}},\
  }\href@noop {} {\  (\bibinfo {year} {2018})},\ \Eprint
  {http://arxiv.org/abs/1810.13423} {arXiv:1810.13423 [astro-ph.CO]}
  \BibitemShut {NoStop}%
\bibitem [{\citenamefont {M\"unchmeyer}\ \emph {et~al.}(2019)\citenamefont
  {M\"unchmeyer}, \citenamefont {Madhavacheril}, \citenamefont {Ferraro},
  \citenamefont {Johnson},\ and\ \citenamefont {Smith}}]{Munchmeyer:2018eey}%
  \BibitemOpen
  \bibfield  {author} {\bibinfo {author} {\bibfnamefont {M.}~\bibnamefont
  {M\"unchmeyer}}, \bibinfo {author} {\bibfnamefont {M.~S.}\ \bibnamefont
  {Madhavacheril}}, \bibinfo {author} {\bibfnamefont {S.}~\bibnamefont
  {Ferraro}}, \bibinfo {author} {\bibfnamefont {M.~C.}\ \bibnamefont
  {Johnson}}, \ and\ \bibinfo {author} {\bibfnamefont {K.~M.}\ \bibnamefont
  {Smith}},\ }\href {\doibase 10.1103/PhysRevD.100.083508} {\bibfield
  {journal} {\bibinfo  {journal} {Phys. Rev. D}\ }\textbf {\bibinfo {volume}
  {100}},\ \bibinfo {pages} {083508} (\bibinfo {year} {2019})},\ \Eprint
  {http://arxiv.org/abs/1810.13424} {arXiv:1810.13424 [astro-ph.CO]}
  \BibitemShut {NoStop}%
\bibitem [{\citenamefont {Zhang}\ and\ \citenamefont
  {Johnson}(2015)}]{Zhang:2015uta}%
  \BibitemOpen
  \bibfield  {author} {\bibinfo {author} {\bibfnamefont {P.}~\bibnamefont
  {Zhang}}\ and\ \bibinfo {author} {\bibfnamefont {M.~C.}\ \bibnamefont
  {Johnson}},\ }\href {\doibase 10.1088/1475-7516/2015/06/046} {\bibfield
  {journal} {\bibinfo  {journal} {JCAP}\ }\textbf {\bibinfo {volume} {06}},\
  \bibinfo {pages} {046} (\bibinfo {year} {2015})},\ \Eprint
  {http://arxiv.org/abs/1501.00511} {arXiv:1501.00511 [astro-ph.CO]}
  \BibitemShut {NoStop}%
\bibitem [{\citenamefont {Hotinli}\ \emph {et~al.}(2019)\citenamefont
  {Hotinli}, \citenamefont {Mertens}, \citenamefont {Johnson},\ and\
  \citenamefont {Kamionkowski}}]{Hotinli:2019wdp}%
  \BibitemOpen
  \bibfield  {author} {\bibinfo {author} {\bibfnamefont {S.~C.}\ \bibnamefont
  {Hotinli}}, \bibinfo {author} {\bibfnamefont {J.~B.}\ \bibnamefont
  {Mertens}}, \bibinfo {author} {\bibfnamefont {M.~C.}\ \bibnamefont
  {Johnson}}, \ and\ \bibinfo {author} {\bibfnamefont {M.}~\bibnamefont
  {Kamionkowski}},\ }\href {\doibase 10.1103/PhysRevD.100.103528} {\bibfield
  {journal} {\bibinfo  {journal} {Phys. Rev. D}\ }\textbf {\bibinfo {volume}
  {100}},\ \bibinfo {pages} {103528} (\bibinfo {year} {2019})},\ \Eprint
  {http://arxiv.org/abs/1908.08953} {arXiv:1908.08953 [astro-ph.CO]}
  \BibitemShut {NoStop}%
\bibitem [{\citenamefont {Cayuso}\ and\ \citenamefont
  {Johnson}(2020)}]{Cayuso:2019hen}%
  \BibitemOpen
  \bibfield  {author} {\bibinfo {author} {\bibfnamefont {J.~I.}\ \bibnamefont
  {Cayuso}}\ and\ \bibinfo {author} {\bibfnamefont {M.~C.}\ \bibnamefont
  {Johnson}},\ }\href {\doibase 10.1103/PhysRevD.101.123508} {\bibfield
  {journal} {\bibinfo  {journal} {Phys. Rev. D}\ }\textbf {\bibinfo {volume}
  {101}},\ \bibinfo {pages} {123508} (\bibinfo {year} {2020})},\ \Eprint
  {http://arxiv.org/abs/1904.10981} {arXiv:1904.10981 [astro-ph.CO]}
  \BibitemShut {NoStop}%
\bibitem [{\citenamefont {Alvarez}\ \emph {et~al.}(2020)\citenamefont
  {Alvarez}, \citenamefont {Ferraro}, \citenamefont {Hill}, \citenamefont
  {Hlo\v{z}ek},\ and\ \citenamefont {Ikape}}]{Alvarez:2020gvl}%
  \BibitemOpen
  \bibfield  {author} {\bibinfo {author} {\bibfnamefont {M.~A.}\ \bibnamefont
  {Alvarez}}, \bibinfo {author} {\bibfnamefont {S.}~\bibnamefont {Ferraro}},
  \bibinfo {author} {\bibfnamefont {J.~C.}\ \bibnamefont {Hill}}, \bibinfo
  {author} {\bibfnamefont {R.}~\bibnamefont {Hlo\v{z}ek}}, \ and\ \bibinfo
  {author} {\bibfnamefont {M.}~\bibnamefont {Ikape}},\ }\href@noop {} {\
  (\bibinfo {year} {2020})},\ \Eprint {http://arxiv.org/abs/2006.06594}
  {arXiv:2006.06594 [astro-ph.CO]} \BibitemShut {NoStop}%
\bibitem [{\citenamefont {Ferraro}\ and\ \citenamefont
  {Smith}(2018)}]{Ferraro:2018izc}%
  \BibitemOpen
  \bibfield  {author} {\bibinfo {author} {\bibfnamefont {S.}~\bibnamefont
  {Ferraro}}\ and\ \bibinfo {author} {\bibfnamefont {K.~M.}\ \bibnamefont
  {Smith}},\ }\href {\doibase 10.1103/PhysRevD.98.123519} {\bibfield  {journal}
  {\bibinfo  {journal} {Phys. Rev. D}\ }\textbf {\bibinfo {volume} {98}},\
  \bibinfo {pages} {123519} (\bibinfo {year} {2018})},\ \Eprint
  {http://arxiv.org/abs/1803.07036} {arXiv:1803.07036 [astro-ph.CO]}
  \BibitemShut {NoStop}%
\bibitem [{\citenamefont {Smith}\ and\ \citenamefont
  {Ferraro}(2017)}]{Smith:2016lnt}%
  \BibitemOpen
  \bibfield  {author} {\bibinfo {author} {\bibfnamefont {K.~M.}\ \bibnamefont
  {Smith}}\ and\ \bibinfo {author} {\bibfnamefont {S.}~\bibnamefont
  {Ferraro}},\ }\href {\doibase 10.1103/PhysRevLett.119.021301} {\bibfield
  {journal} {\bibinfo  {journal} {Phys. Rev. Lett.}\ }\textbf {\bibinfo
  {volume} {119}},\ \bibinfo {pages} {021301} (\bibinfo {year} {2017})},\
  \Eprint {http://arxiv.org/abs/1607.01769} {arXiv:1607.01769 [astro-ph.CO]}
  \BibitemShut {NoStop}%
\bibitem [{\citenamefont {Hotinli}\ and\ \citenamefont
  {Johnson}(2020)}]{Hotinli:2020csk}%
  \BibitemOpen
  \bibfield  {author} {\bibinfo {author} {\bibfnamefont {S.~C.}\ \bibnamefont
  {Hotinli}}\ and\ \bibinfo {author} {\bibfnamefont {M.~C.}\ \bibnamefont
  {Johnson}},\ }\href@noop {} {\  (\bibinfo {year} {2020})},\ \Eprint
  {http://arxiv.org/abs/2012.09851} {arXiv:2012.09851 [astro-ph.CO]}
  \BibitemShut {NoStop}%
\bibitem [{\citenamefont {Aghamousa}\ \emph {et~al.}(2016)\citenamefont
  {Aghamousa} \emph {et~al.}}]{DESI:2016fyo}%
  \BibitemOpen
  \bibfield  {author} {\bibinfo {author} {\bibfnamefont {A.}~\bibnamefont
  {Aghamousa}} \emph {et~al.} (\bibinfo {collaboration} {DESI}),\ }\href@noop
  {} {\  (\bibinfo {year} {2016})},\ \Eprint {http://arxiv.org/abs/1611.00036}
  {arXiv:1611.00036 [astro-ph.IM]} \BibitemShut {NoStop}%
\bibitem [{\citenamefont {{LSST Science Collaboration}}\ \emph
  {et~al.}(2009)\citenamefont {{LSST Science Collaboration}} \emph
  {et~al.}}]{2009arXiv0912.0201L}%
  \BibitemOpen
  \bibfield  {author} {\bibinfo {author} {\bibnamefont {{LSST Science
  Collaboration}}} \emph {et~al.},\ }\href@noop {} {\bibfield  {journal}
  {\bibinfo  {journal} {ArXiv e-prints}\ } (\bibinfo {year} {2009})},\ \Eprint
  {http://arxiv.org/abs/0912.0201} {arXiv:0912.0201 [astro-ph.IM]} \BibitemShut
  {NoStop}%
\bibitem [{\citenamefont {Abazajian}\ \emph
  {et~al.}(2016{\natexlab{a}})\citenamefont {Abazajian} \emph
  {et~al.}}]{Abazajian:2016yjj}%
  \BibitemOpen
  \bibfield  {author} {\bibinfo {author} {\bibfnamefont {K.~N.}\ \bibnamefont
  {Abazajian}} \emph {et~al.} (\bibinfo {collaboration} {CMB-S4}),\ }\href@noop
  {} {\  (\bibinfo {year} {2016}{\natexlab{a}})},\ \Eprint
  {http://arxiv.org/abs/1610.02743} {arXiv:1610.02743 [astro-ph.CO]}
  \BibitemShut {NoStop}%
%%CITATION = ARXIV:1610.02743;%%
\bibitem [{\citenamefont {Abazajian}\ \emph {et~al.}(2019)\citenamefont
  {Abazajian} \emph {et~al.}}]{Abazajian:2019eic}%
  \BibitemOpen
  \bibfield  {author} {\bibinfo {author} {\bibfnamefont {K.}~\bibnamefont
  {Abazajian}} \emph {et~al.},\ }\href@noop {} {\  (\bibinfo {year} {2019})},\
  \Eprint {http://arxiv.org/abs/1907.04473} {arXiv:1907.04473 [astro-ph.IM]}
  \BibitemShut {NoStop}%
\bibitem [{\citenamefont {Schlegel}\ \emph {et~al.}(2019)\citenamefont
  {Schlegel} \emph {et~al.}}]{Schlegel:2019eqc}%
  \BibitemOpen
  \bibfield  {author} {\bibinfo {author} {\bibfnamefont {D.~J.}\ \bibnamefont
  {Schlegel}} \emph {et~al.},\ }\href@noop {} {\  (\bibinfo {year} {2019})},\
  \Eprint {http://arxiv.org/abs/1907.11171} {arXiv:1907.11171 [astro-ph.IM]}
  \BibitemShut {NoStop}%
\bibitem [{\citenamefont {Ferraro}\ \emph {et~al.}(2022)\citenamefont
  {Ferraro}, \citenamefont {Sailer}, \citenamefont {Slosar},\ and\
  \citenamefont {White}}]{Ferraro:2022cmj}%
  \BibitemOpen
  \bibfield  {author} {\bibinfo {author} {\bibfnamefont {S.}~\bibnamefont
  {Ferraro}}, \bibinfo {author} {\bibfnamefont {N.}~\bibnamefont {Sailer}},
  \bibinfo {author} {\bibfnamefont {A.}~\bibnamefont {Slosar}}, \ and\ \bibinfo
  {author} {\bibfnamefont {M.}~\bibnamefont {White}},\ }\href@noop {} {\
  (\bibinfo {year} {2022})},\ \Eprint {http://arxiv.org/abs/2203.07506}
  {arXiv:2203.07506 [astro-ph.CO]} \BibitemShut {NoStop}%
\bibitem [{\citenamefont {Ali-Haimoud}\ and\ \citenamefont
  {Hirata}(2011)}]{Ali-Haimoud:2010hou}%
  \BibitemOpen
  \bibfield  {author} {\bibinfo {author} {\bibfnamefont {Y.}~\bibnamefont
  {Ali-Haimoud}}\ and\ \bibinfo {author} {\bibfnamefont {C.~M.}\ \bibnamefont
  {Hirata}},\ }\href {\doibase 10.1103/PhysRevD.83.043513} {\bibfield
  {journal} {\bibinfo  {journal} {Phys. Rev. D}\ }\textbf {\bibinfo {volume}
  {83}},\ \bibinfo {pages} {043513} (\bibinfo {year} {2011})},\ \Eprint
  {http://arxiv.org/abs/1011.3758} {arXiv:1011.3758 [astro-ph.CO]} \BibitemShut
  {NoStop}%
\bibitem [{\citenamefont {La~Plante}\ \emph {et~al.}(2017)\citenamefont
  {La~Plante}, \citenamefont {Trac}, \citenamefont {Croft},\ and\ \citenamefont
  {Cen}}]{LaPlante:2016bzu}%
  \BibitemOpen
  \bibfield  {author} {\bibinfo {author} {\bibfnamefont {P.}~\bibnamefont
  {La~Plante}}, \bibinfo {author} {\bibfnamefont {H.}~\bibnamefont {Trac}},
  \bibinfo {author} {\bibfnamefont {R.}~\bibnamefont {Croft}}, \ and\ \bibinfo
  {author} {\bibfnamefont {R.}~\bibnamefont {Cen}},\ }\href {\doibase
  10.3847/1538-4357/aa7136} {\bibfield  {journal} {\bibinfo  {journal}
  {Astrophys. J.}\ }\textbf {\bibinfo {volume} {841}},\ \bibinfo {pages} {87}
  (\bibinfo {year} {2017})},\ \Eprint {http://arxiv.org/abs/1610.02047}
  {arXiv:1610.02047 [astro-ph.CO]} \BibitemShut {NoStop}%
\bibitem [{\citenamefont {{Lusso}}\ \emph {et~al.}(2015)\citenamefont
  {{Lusso}}, \citenamefont {{Worseck}}, \citenamefont {{Hennawi}},
  \citenamefont {{Prochaska}}, \citenamefont {{Vignali}}, \citenamefont
  {{Stern}},\ and\ \citenamefont {{O'Meara}}}]{Lusso2015}%
  \BibitemOpen
  \bibfield  {author} {\bibinfo {author} {\bibfnamefont {E.}~\bibnamefont
  {{Lusso}}}, \bibinfo {author} {\bibfnamefont {G.}~\bibnamefont {{Worseck}}},
  \bibinfo {author} {\bibfnamefont {J.~F.}\ \bibnamefont {{Hennawi}}}, \bibinfo
  {author} {\bibfnamefont {J.~X.}\ \bibnamefont {{Prochaska}}}, \bibinfo
  {author} {\bibfnamefont {C.}~\bibnamefont {{Vignali}}}, \bibinfo {author}
  {\bibfnamefont {J.}~\bibnamefont {{Stern}}}, \ and\ \bibinfo {author}
  {\bibfnamefont {J.~M.}\ \bibnamefont {{O'Meara}}},\ }\href {\doibase
  10.1093/mnras/stv516} {\bibfield  {journal} {\bibinfo  {journal} {MNRAS}\
  }\textbf {\bibinfo {volume} {449}},\ \bibinfo {pages} {4204} (\bibinfo {year}
  {2015})},\ \Eprint {http://arxiv.org/abs/1503.02075} {arXiv:1503.02075
  [astro-ph.GA]} \BibitemShut {NoStop}%
\bibitem [{\citenamefont {{White}}\ \emph {et~al.}(2012)\citenamefont
  {{White}}, \citenamefont {{Myers}}, \citenamefont {{Ross}}, \citenamefont
  {{Schlegel}}, \citenamefont {{Hennawi}}, \citenamefont {{Shen}},
  \citenamefont {{McGreer}}, \citenamefont {{Strauss}}, \citenamefont
  {{Bolton}}, \citenamefont {{Bovy}}, \citenamefont {{Fan}}, \citenamefont
  {{Miralda-Escude}}, \citenamefont {{Palanque-Delabrouille}}, \citenamefont
  {{Paris}}, \citenamefont {{Petitjean}}, \citenamefont {{Schneider}},
  \citenamefont {{Viel}}, \citenamefont {{Weinberg}}, \citenamefont {{Yeche}},
  \citenamefont {{Zehavi}}, \citenamefont {{Pan}}, \citenamefont {{Snedden}},
  \citenamefont {{Bizyaev}}, \citenamefont {{Brewington}}, \citenamefont
  {{Brinkmann}}, \citenamefont {{Malanushenko}}, \citenamefont
  {{Malanushenko}}, \citenamefont {{Oravetz}}, \citenamefont {{Simmons}},
  \citenamefont {{Sheldon}},\ and\ \citenamefont {{Weaver}}}]{White2012}%
  \BibitemOpen
  \bibfield  {author} {\bibinfo {author} {\bibfnamefont {M.}~\bibnamefont
  {{White}}}, \bibinfo {author} {\bibfnamefont {A.~D.}\ \bibnamefont
  {{Myers}}}, \bibinfo {author} {\bibfnamefont {N.~P.}\ \bibnamefont {{Ross}}},
  \bibinfo {author} {\bibfnamefont {D.~J.}\ \bibnamefont {{Schlegel}}},
  \bibinfo {author} {\bibfnamefont {J.~F.}\ \bibnamefont {{Hennawi}}}, \bibinfo
  {author} {\bibfnamefont {Y.}~\bibnamefont {{Shen}}}, \bibinfo {author}
  {\bibfnamefont {I.}~\bibnamefont {{McGreer}}}, \bibinfo {author}
  {\bibfnamefont {M.~A.}\ \bibnamefont {{Strauss}}}, \bibinfo {author}
  {\bibfnamefont {A.~S.}\ \bibnamefont {{Bolton}}}, \bibinfo {author}
  {\bibfnamefont {J.}~\bibnamefont {{Bovy}}}, \bibinfo {author} {\bibfnamefont
  {X.}~\bibnamefont {{Fan}}}, \bibinfo {author} {\bibfnamefont
  {J.}~\bibnamefont {{Miralda-Escude}}}, \bibinfo {author} {\bibfnamefont
  {N.}~\bibnamefont {{Palanque-Delabrouille}}}, \bibinfo {author}
  {\bibfnamefont {I.}~\bibnamefont {{Paris}}}, \bibinfo {author} {\bibfnamefont
  {P.}~\bibnamefont {{Petitjean}}}, \bibinfo {author} {\bibfnamefont {D.~P.}\
  \bibnamefont {{Schneider}}}, \bibinfo {author} {\bibfnamefont
  {M.}~\bibnamefont {{Viel}}}, \bibinfo {author} {\bibfnamefont {D.~H.}\
  \bibnamefont {{Weinberg}}}, \bibinfo {author} {\bibfnamefont
  {C.}~\bibnamefont {{Yeche}}}, \bibinfo {author} {\bibfnamefont
  {I.}~\bibnamefont {{Zehavi}}}, \bibinfo {author} {\bibfnamefont
  {K.}~\bibnamefont {{Pan}}}, \bibinfo {author} {\bibfnamefont
  {S.}~\bibnamefont {{Snedden}}}, \bibinfo {author} {\bibfnamefont
  {D.}~\bibnamefont {{Bizyaev}}}, \bibinfo {author} {\bibfnamefont
  {H.}~\bibnamefont {{Brewington}}}, \bibinfo {author} {\bibfnamefont
  {J.}~\bibnamefont {{Brinkmann}}}, \bibinfo {author} {\bibfnamefont
  {V.}~\bibnamefont {{Malanushenko}}}, \bibinfo {author} {\bibfnamefont
  {E.}~\bibnamefont {{Malanushenko}}}, \bibinfo {author} {\bibfnamefont
  {D.}~\bibnamefont {{Oravetz}}}, \bibinfo {author} {\bibfnamefont
  {A.}~\bibnamefont {{Simmons}}}, \bibinfo {author} {\bibfnamefont
  {A.}~\bibnamefont {{Sheldon}}}, \ and\ \bibinfo {author} {\bibfnamefont
  {B.~A.}\ \bibnamefont {{Weaver}}},\ }\href {\doibase
  10.1111/j.1365-2966.2012.21251.x} {\bibfield  {journal} {\bibinfo  {journal}
  {MNRAS}\ }\textbf {\bibinfo {volume} {424}},\ \bibinfo {pages} {933}
  (\bibinfo {year} {2012})},\ \Eprint {http://arxiv.org/abs/1203.5306}
  {arXiv:1203.5306 [astro-ph.CO]} \BibitemShut {NoStop}%
\bibitem [{\citenamefont {{Haardt}}\ and\ \citenamefont
  {{Madau}}(2012)}]{Haardt2012}%
  \BibitemOpen
  \bibfield  {author} {\bibinfo {author} {\bibfnamefont {F.}~\bibnamefont
  {{Haardt}}}\ and\ \bibinfo {author} {\bibfnamefont {P.}~\bibnamefont
  {{Madau}}},\ }\href {\doibase 10.1088/0004-637X/746/2/125} {\bibfield
  {journal} {\bibinfo  {journal} {\apj}\ }\textbf {\bibinfo {volume} {746}},\
  \bibinfo {eid} {125} (\bibinfo {year} {2012})},\ \Eprint
  {http://arxiv.org/abs/1105.2039} {arXiv:1105.2039 [astro-ph.CO]} \BibitemShut
  {NoStop}%
\bibitem [{\citenamefont {Foreman}\ \emph {et~al.}(2022)\citenamefont
  {Foreman}, \citenamefont {Hotinli} \emph {et~al.}}]{Foreman:2022ksz}%
  \BibitemOpen
  \bibfield  {author} {\bibinfo {author} {\bibfnamefont {S.}~\bibnamefont
  {Foreman}}, \bibinfo {author} {\bibfnamefont {S.~C.}\ \bibnamefont
  {Hotinli}},  \emph {et~al.},\ }\href@noop {} {\  (\bibinfo {year}
  {2022})}\BibitemShut {NoStop}%
\bibitem [{\citenamefont {Ferraro}\ \emph {et~al.}(2019)\citenamefont {Ferraro}
  \emph {et~al.}}]{Ferraro:2019uce}%
  \BibitemOpen
  \bibfield  {author} {\bibinfo {author} {\bibfnamefont {S.}~\bibnamefont
  {Ferraro}} \emph {et~al.},\ }\href@noop {} {\  (\bibinfo {year} {2019})},\
  \Eprint {http://arxiv.org/abs/1903.09208} {arXiv:1903.09208 [astro-ph.CO]}
  \BibitemShut {NoStop}%
\bibitem [{\citenamefont {Harikane}\ \emph {et~al.}(2017)\citenamefont
  {Harikane} \emph {et~al.}}]{Harikane:2017lcw}%
  \BibitemOpen
  \bibfield  {author} {\bibinfo {author} {\bibfnamefont {Y.}~\bibnamefont
  {Harikane}} \emph {et~al.},\ }\href {\doibase 10.1093/pasj/psx097} {\
  (\bibinfo {year} {2017}),\ 10.1093/pasj/psx097},\ \Eprint
  {http://arxiv.org/abs/1704.06535} {arXiv:1704.06535 [astro-ph.GA]}
  \BibitemShut {NoStop}%
\bibitem [{\citenamefont {{DESI Collaboration}}(2016)}]{2016arXiv161100036D}%
  \BibitemOpen
  \bibfield  {author} {\bibinfo {author} {\bibnamefont {{DESI
  Collaboration}}},\ }\href@noop {} {\bibfield  {journal} {\bibinfo  {journal}
  {arXiv e-prints}\ ,\ \bibinfo {eid} {arXiv:1611.00036}} (\bibinfo {year}
  {2016})},\ \Eprint {http://arxiv.org/abs/1611.00036} {arXiv:1611.00036
  [astro-ph.IM]} \BibitemShut {NoStop}%
\bibitem [{\citenamefont {Hotinli}\ \emph {et~al.}(2021)\citenamefont
  {Hotinli}, \citenamefont {Smith}, \citenamefont {Madhavacheril},\ and\
  \citenamefont {Kamionkowski}}]{Hotinli:2021hih}%
  \BibitemOpen
  \bibfield  {author} {\bibinfo {author} {\bibfnamefont {S.~C.}\ \bibnamefont
  {Hotinli}}, \bibinfo {author} {\bibfnamefont {K.~M.}\ \bibnamefont {Smith}},
  \bibinfo {author} {\bibfnamefont {M.~S.}\ \bibnamefont {Madhavacheril}}, \
  and\ \bibinfo {author} {\bibfnamefont {M.}~\bibnamefont {Kamionkowski}},\
  }\href {\doibase 10.1103/PhysRevD.104.083529} {\bibfield  {journal} {\bibinfo
   {journal} {Phys. Rev. D}\ }\textbf {\bibinfo {volume} {104}},\ \bibinfo
  {pages} {083529} (\bibinfo {year} {2021})},\ \Eprint
  {http://arxiv.org/abs/2108.02207} {arXiv:2108.02207 [astro-ph.CO]}
  \BibitemShut {NoStop}%
\bibitem [{\citenamefont {{Lewis}}\ \emph {et~al.}(2000)\citenamefont
  {{Lewis}}, \citenamefont {{Challinor}},\ and\ \citenamefont
  {{Lasenby}}}]{CAMB}%
  \BibitemOpen
  \bibfield  {author} {\bibinfo {author} {\bibfnamefont {A.}~\bibnamefont
  {{Lewis}}}, \bibinfo {author} {\bibfnamefont {A.}~\bibnamefont
  {{Challinor}}}, \ and\ \bibinfo {author} {\bibfnamefont {A.}~\bibnamefont
  {{Lasenby}}},\ }\href {\doibase 10.1086/309179} {\bibfield  {journal}
  {\bibinfo  {journal} {\apj}\ }\textbf {\bibinfo {volume} {538}},\ \bibinfo
  {pages} {473} (\bibinfo {year} {2000})},\ \Eprint
  {http://arxiv.org/abs/astro-ph/9911177} {arXiv:astro-ph/9911177 [astro-ph]}
  \BibitemShut {NoStop}%
\bibitem [{\citenamefont {Aver}\ \emph {et~al.}(2015)\citenamefont {Aver},
  \citenamefont {Olive},\ and\ \citenamefont {Skillman}}]{Aver:2015iza}%
  \BibitemOpen
  \bibfield  {author} {\bibinfo {author} {\bibfnamefont {E.}~\bibnamefont
  {Aver}}, \bibinfo {author} {\bibfnamefont {K.~A.}\ \bibnamefont {Olive}}, \
  and\ \bibinfo {author} {\bibfnamefont {E.~D.}\ \bibnamefont {Skillman}},\
  }\href {\doibase 10.1088/1475-7516/2015/07/011} {\bibfield  {journal}
  {\bibinfo  {journal} {JCAP}\ }\textbf {\bibinfo {volume} {07}},\ \bibinfo
  {pages} {011} (\bibinfo {year} {2015})},\ \Eprint
  {http://arxiv.org/abs/1503.08146} {arXiv:1503.08146 [astro-ph.CO]}
  \BibitemShut {NoStop}%
\bibitem [{\citenamefont {Aghanim}\ \emph {et~al.}(2020)\citenamefont {Aghanim}
  \emph {et~al.}}]{Planck:2018vyg}%
  \BibitemOpen
  \bibfield  {author} {\bibinfo {author} {\bibfnamefont {N.}~\bibnamefont
  {Aghanim}} \emph {et~al.} (\bibinfo {collaboration} {Planck}),\ }\href
  {\doibase 10.1051/0004-6361/201833910} {\bibfield  {journal} {\bibinfo
  {journal} {Astron. Astrophys.}\ }\textbf {\bibinfo {volume} {641}},\ \bibinfo
  {pages} {A6} (\bibinfo {year} {2020})},\ \bibinfo {note} {[Erratum:
  Astron.Astrophys. 652, C4 (2021)]},\ \Eprint
  {http://arxiv.org/abs/1807.06209} {arXiv:1807.06209 [astro-ph.CO]}
  \BibitemShut {NoStop}%
\bibitem [{\citenamefont {Abazajian}\ \emph
  {et~al.}(2016{\natexlab{b}})\citenamefont {Abazajian} \emph
  {et~al.}}]{CMB-S4:2016ple}%
  \BibitemOpen
  \bibfield  {author} {\bibinfo {author} {\bibfnamefont {K.~N.}\ \bibnamefont
  {Abazajian}} \emph {et~al.} (\bibinfo {collaboration} {CMB-S4}),\ }\href@noop
  {} {\  (\bibinfo {year} {2016}{\natexlab{b}})},\ \Eprint
  {http://arxiv.org/abs/1610.02743} {arXiv:1610.02743 [astro-ph.CO]}
  \BibitemShut {NoStop}%
\bibitem [{\citenamefont {Hotinli}\ \emph {et~al.}(2022)\citenamefont
  {Hotinli}, \citenamefont {Meyers}, \citenamefont {Trendafilova},
  \citenamefont {Green},\ and\ \citenamefont {van Engelen}}]{Hotinli:2021umk}%
  \BibitemOpen
  \bibfield  {author} {\bibinfo {author} {\bibfnamefont {S.~C.}\ \bibnamefont
  {Hotinli}}, \bibinfo {author} {\bibfnamefont {J.}~\bibnamefont {Meyers}},
  \bibinfo {author} {\bibfnamefont {C.}~\bibnamefont {Trendafilova}}, \bibinfo
  {author} {\bibfnamefont {D.}~\bibnamefont {Green}}, \ and\ \bibinfo {author}
  {\bibfnamefont {A.}~\bibnamefont {van Engelen}},\ }\href {\doibase
  10.1088/1475-7516/2022/04/020} {\bibfield  {journal} {\bibinfo  {journal}
  {JCAP}\ }\textbf {\bibinfo {volume} {04}},\ \bibinfo {pages} {020} (\bibinfo
  {year} {2022})},\ \Eprint {http://arxiv.org/abs/2111.15036} {arXiv:2111.15036
  [astro-ph.CO]} \BibitemShut {NoStop}%
\bibitem [{\citenamefont {Cayuso}\ \emph {et~al.}()\citenamefont {Cayuso},
  \citenamefont {Bloch}, \citenamefont {Hotinli}, \citenamefont {Johnson},\
  and\ \citenamefont {McCarthy}}]{Cayuso:2021ljq}%
  \BibitemOpen
  \bibfield  {author} {\bibinfo {author} {\bibfnamefont {J.}~\bibnamefont
  {Cayuso}}, \bibinfo {author} {\bibfnamefont {R.}~\bibnamefont {Bloch}},
  \bibinfo {author} {\bibfnamefont {S.~C.}\ \bibnamefont {Hotinli}}, \bibinfo
  {author} {\bibfnamefont {M.~C.}\ \bibnamefont {Johnson}}, \ and\ \bibinfo
  {author} {\bibfnamefont {F.}~\bibnamefont {McCarthy}},\ }\href@noop {} {\
  }\Eprint {http://arxiv.org/abs/2111.11526} {arXiv:2111.11526 [astro-ph.CO]}
  \BibitemShut {NoStop}%
\bibitem [{\citenamefont {{Deutsch}}\ \emph {et~al.}(2018)\citenamefont
  {{Deutsch}}, \citenamefont {{Johnson}}, \citenamefont {{M{\"u}nchmeyer}},\
  and\ \citenamefont {{Terrana}}}]{Deutsch:2017cja}%
  \BibitemOpen
  \bibfield  {author} {\bibinfo {author} {\bibfnamefont {A.-S.}\ \bibnamefont
  {{Deutsch}}}, \bibinfo {author} {\bibfnamefont {M.~C.}\ \bibnamefont
  {{Johnson}}}, \bibinfo {author} {\bibfnamefont {M.}~\bibnamefont
  {{M{\"u}nchmeyer}}}, \ and\ \bibinfo {author} {\bibfnamefont
  {A.}~\bibnamefont {{Terrana}}},\ }\href {\doibase
  10.1088/1475-7516/2018/04/034} {\bibfield  {journal} {\bibinfo  {journal}
  {JCAP}\ }\textbf {\bibinfo {volume} {2018}},\ \bibinfo {eid} {034} (\bibinfo
  {year} {2018})},\ \Eprint {http://arxiv.org/abs/1705.08907} {arXiv:1705.08907
  [astro-ph.CO]} \BibitemShut {NoStop}%
\end{thebibliography}%
\end{document}